%% file: DDSDI.tex
\documentclass[twocolumn,aps,prx,showpacs,amsmath,superscriptaddress,longbibliography,notitlepage]{revtex4-2}
\usepackage{amssymb}
\usepackage{mathrsfs}
\usepackage{graphicx}
\usepackage{float}
\usepackage[caption=false]{subfig}
\usepackage[pdftex,colorlinks,linkcolor={red!75!black},citecolor={red!75!black},urlcolor={blue!75!black}]{hyperref}
\usepackage{verbatim}
\usepackage{epstopdf}
\usepackage[normalem]{ulem}
\usepackage{verbatim}
\usepackage{xcolor}
\usepackage{bm}
\renewcommand{\Re}{\operatorname{Re}}
\renewcommand{\Im}{\operatorname{Im}}
\newcommand{\clr}{\color{red!75!black}}

\newcommand{\Rnum}[1]{\uppercase\expandafter{\romannumeral #1\relax}}

\begin{document}
	
	\title{Dynamical Degeneracy Splitting and Directional Invisibility in Non-Hermitian Systems}
	
	\author{Kai Zhang}
	\affiliation{Beijing National Laboratory for Condensed Matter Physics, and Institute of Physics, Chinese Academy of Sciences, Beijing 100190, China}
	\author{Chen Fang}
	\email{cfang@iphy.ac.cn}
	\affiliation{Beijing National Laboratory for Condensed Matter Physics, and Institute of Physics, Chinese Academy of Sciences, Beijing 100190, China}
	\affiliation{Songshan Lake Materials Laboratory, Dongguan, Guangdong 523808, China}
	\affiliation{Kavli Institute for Theoretical Sciences, Chinese Academy of Sciences, Beijing 100190, China}
	\author{Zhesen Yang}
	\email{yangzs@ucas.ac.cn}
	\affiliation{Kavli Institute for Theoretical Sciences, Chinese Academy of Sciences, Beijing 100190, China}
	
\begin{abstract}
In this paper, we introduce the concept of dynamical degeneracy splitting to describe the anisotropic decay behaviors in non-Hermitian systems. 
We demonstrate that systems with dynamical degeneracy splitting exhibit two distinctive features: 
(i) the system shows frequency-resolved non-Hermitian skin effect; 
(ii) Green's function exhibits anomalous at given frequency, leading to uneven broadening in spectral function and anomalous scattering. 
As an application, we propose directional invisibility based on wave packet dynamics to investigate the geometry-dependent skin effect in higher dimensions. 
Our work elucidates a faithful correspondence between non-Hermitian skin effect and Green’s function, offering a guiding principle for exploration of novel physical phenomena emerging from this effect. 
\end{abstract}

\maketitle

\emph{{\clr Introduction}.---}~Non-Hermitian Hamiltonians~\cite{Rotter2009,Moiseyev2011,Brody2014,Ashida2020,Bergholtz2021_RMP,DingKun2022_NRP}, which can effectively capture the dynamics of the system that is coupled to external environments, have been implemented in a wide variety of systems~\cite{Regensburger2012,Gao2015_Nature,FengLiang2017,Miri2019,YangLan2019,Vincenzo2020NP,Vincenzo2020,DiZhou2020,FuLiang2017_arXiv,ShenHT2018_PRL,FuLiang2020_PRL,McDonald2018_PRX}.
Without the constraint of Hermiticity, the eigenvalues of the Hamiltonian can be complex, leading to many intriguing phenomena in non-Hermitian systems. 
One such phenomenon is the non-Hermitian skin effect (NHSE)~\cite{Yao2018,Kunst2018_PRL,WangZhong2018,Torres2018,Murakami2019_PRL,SongFei2019,Sato2019_PRX,ChingHua2019,LeeCH2019_PRL,Ashvin2019_PRL,Nori2019,DengTS2019,Longhi2019_PRR,Kai2020,Slager2020PRL,Okuma2020_PRL,Zhesen2020_aGBZ,Zhesen2020_SE,Titus2020,NeupertETI2020,Kawabata2020,LiLinhuPRL2020,LiLH2020_NC,ChuCH2020_PRR,Kawabata2020_Symplectic,Thomale2020,Ghatak2020,XuePeng2020,XueWT2021_PRB,SunXQ2021_PRL,Kawabata2021_PRL,Sato2021_PRL,LiLH2021_NC,LiuYX2021_PRB,GuoCX2021_PRL,LuMing2021,MaoLiang2021,SongFei2022,Kai2022NC,DingKun2022_NanoPho,Wang2022_NC,Longhi2022_PRL,Longhi2022_PRB,LiangQ2022_PRL,ChenWei2022_arXiv}, which refers to that extensive eigenstates of system-size order are concentrated on the open boundaries; and that the energy spectrum is highly sensitive to the change of boundary conditions. 
In one dimension, the NHSE can be well understood within the framework of the non-Bloch band theory~\cite{Yao2018,Kunst2018_PRL,Murakami2019_PRL,Zhesen2020_aGBZ}. 
Extending to higher dimensions, a bunch of new appearances in NHSE have been discovered~\cite{LeeCH2019_PRL,Kawabata2020,Kai2022NC,SongFei2022} and the generalization of the non-Bloch theory has also been attempted~\cite{WangZhong2018,HuiJiang2022,Murakami2022,HYWang2022}. 
However, up to date, many questions still need to be solved. 
One representative is the recently discovered geometry-dependent skin effect~\cite{Kai2022NC}, where the appearance of NHSE clearly depends on the geometric shapes of the open boundary. 
How to interpret and find a guiding principle to detect such kind of phenomena in higher dimensions? This is one motivation for this work.  

\begin{figure*}[t]
	\begin{center}
		\includegraphics[width=1\linewidth]{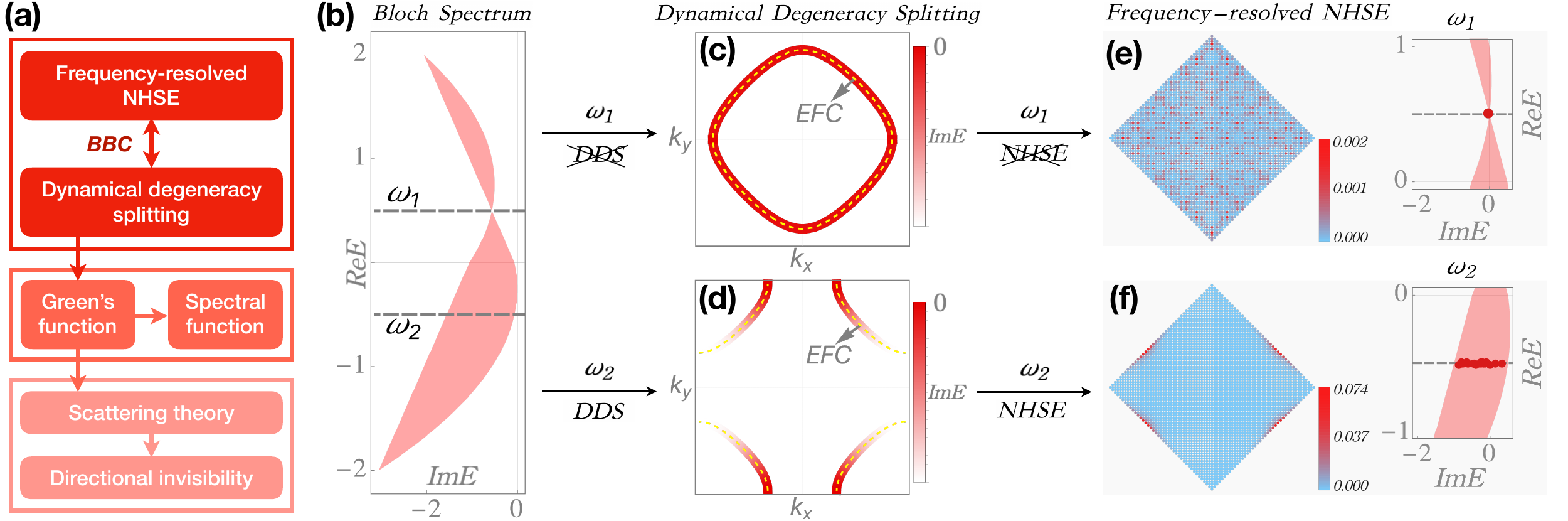}
		\par\end{center}
	\protect\caption{\label{fig:1} 
	(a) Schematic of relation between dynamical degeneracy splitting (DDS), frequency-resolved NHSE, and the physical consequences via Green's function. 
	(b)-(f) show the bulk-boundary correspondence between dynamical degeneracy splitting and frequency-resolved NHSE. 
	For the Bloch spectrum in (b), there is no dynamical degeneracy splitting at excitation frequency $\omega_1$ in (c), where the color bar corresponds to imaginary energy $\mathrm{Im}E(\bm{k})$ on the equal-frequency contour (the dashed yellow lines). 
	Correspondingly, NHSE is absent at this frequency in (e). 
	Dynamical degeneracy splitting occurs at $\omega_2$ as shown in (d), and consequently, NHSE appears at frequency $\omega_{2}$ in (f). } 
\end{figure*}

Another motivation comes from the following considerations. 
Previous studies have suggested that the spectral area of the Bloch spectrum can faithfully predict the presence of higher-dimensional NHSE~\cite{Kai2022NC}; however, this criterion losses the information of band structures around a given frequency~$\omega\in \mathbb{R}$~\cite{Zhou2018}, which is primarily concerned in spectroscopic measurements~\cite{LvBQ2019_NRP}. 
Moreover, the criterion fails to relate NHSE to realistic physical observations. 
For example, in condensed matter physics, the transport and response properties of the system are mainly determined by the excitations near the Fermi energy; thus, only NHSE appearing near the Fermi energy does matter, which cannot be inferred from the spectral-area criterion. 
These considerations inspire us to find a \emph{frequency criterion} for the manifestation of NHSE, thereby delivering NHSE to broader physical scenarios. 

In this paper, we demonstrate that the {\em dynamical degeneracy splitting} plays the role of frequency criterion for the appearance of NHSE.
From the perspective of bulk-boundary correspondence, the dynamical degeneracy splitting reflects the bulk characteristic of the NHSE and is independent of boundary conditions, which implies that the dynamical degeneracy splitting maybe more intrinsic than the NHSE itself. 
As illustrated in Fig.~\ref{fig:1}(a), the dynamical degeneracy splitting not only helps us to understand the physical origin of NHSE, but also reveal a deep relation to the Green's function. 
Given this connection, we establish the scattering theory in non-Hermitian systems, and reveal that when the dynamical degeneracy splitting occurs, the scattered waves will be damped away from impurities. 
Applying the scattering theory to geometry-dependent skin effect, we propose a new phenomenon unique to non-Hermitian systems called {\em directional invisibility}, which refers to that the reflected components of the incident wave packet are visible when the impurity line is in several spatial directions but invisible in the remaining directions.  
As an application, directional invisibility can serve as an experimentally feasible method to directly detect the existence of geometry-dependent skin effect without the need for open boundary conditions. 

\emph{{\clr Dynamical degeneracy splitting and frequency-resolved NHSE}.---}~Now we start with the non-Hermitian Bloch band $E_\mu(\bm{k})$ to explain what dynamical degeneracy splitting is. 
For a given excitation frequency $\omega\in\mathbb{R}$ (the real frequency is assumed throughout this paper), the equal-frequency contour $\bm{K}(\omega)$ can be defined as
\begin{equation}\label{MT_EFCDef}
	\bm{K}(\omega\in\mathbb{R})=\bm{K}_1(\omega\in\mathbb{R})\cup...\cup\bm{K}_m(\omega\in\mathbb{R}),
\end{equation}
where $\bm{K}_\mu(\omega\in\mathbb{R})=\{\bm{k}\in {\rm BZ} | \Re E_\mu(\bm{k})=\omega\}$. 
When $\omega$ is chosen to be the chemical potential in electronic systems, $\bm {K}(\omega)$ is nothing but the renormalized Fermi surface. 
An example is illustrated in Fig.~\ref{fig:1}(b)-(d). 
When $\omega=\omega_{1/2}$, the corresponding equal-frequency contour is plotted in Fig.~\ref{fig:1}(c/d) by yellow dashing lines. 
Physically, each point on the equal-frequency contour in  Fig.~\ref{fig:1}(c) corresponds to an excited mode at frequency $\omega_{1}$, and the corresponding group velocity of this mode in real space is along the normal direction at that point on the equal-frequency contour~\cite{SupMat}. 

In the Hermitian case, all excited modes at frequency $\omega$ are degenerate since $\Im E_\mu(\bm{k})=0$ for all $\bm{k}$. 
Under a generic open boundary geometry, the corresponding eigenstate with energy $\omega$ are constructed by the linear superposition of these $\bm{k}\in \bm{K}(\omega)$. 
However, once the non-Hermitian term is introduced, the imaginary part $\Im E_\mu(\bm{k})$ will broaden the equal-frequency contour in complex ways, which in general will split this degeneracy as shown in Fig.~\ref{fig:1}(d). 
As a result, even under the same open boundary geometry, the original linear superposition of Bloch waves is no longer to be the eigenstate, which implies the emergence of NHSE at frequency $\omega$~\cite{SupMat}. 
We refer to the above type of degeneracy splitting as dynamic degeneracy splitting. 
This phenomenon results from differences in the lifetimes of equal-frequency excitation modes, has a dynamical consequence and corresponds to the frequency-resolved NHSE. 

We use the example $H_{\mathrm{NH}}(\bm{k})=\cos{k_x} + \cos{k_y} + i [ (1/2 - \cos{k_x} - \cos{k_y}) \cos{k_x}]- i 9/16$ to demonstrate this point. 
As shown in Fig.~\ref{fig:1}(b), the dynamical degeneracy splitting occurs at $\omega_2=-1/2$ but not $\omega_1=1/2$. 
It implies that these eigenstates $\psi_{i\in\omega_{2}(\omega_{1})}(\bm{r})$ with eigenvalues satisfying $\Re E_{i,\mathrm{OBC}}=\omega_{2}(\omega_{1})$ will (not) show NHSE at this frequency. 
In Fig.~\ref{fig:1}(e)(f), we plot $P_{\omega}(\bm{r}) = \sum_{i\in\omega}|\psi_i(\bm{r})|^2$ on the diamond geometry with lattice size $L_x=L_y=80$. 
It is shown that $P_{\omega_1}(\bm{r})$ is extensive on the entire lattice in (e), but $P_{\omega_2}(\bm{r})$  shows localization behavior in (f), which demonstrates the correspondence between the dynamical degeneracy splitting and frequency-resolved NHSE. 

{\em \clr Dynamical degeneracy splitting in Green's function.}---
Apart from the relation to NHSE, dynamical degeneracy splitting is also associated with Green's function as illustrated in Fig.~\ref{fig:1}(a). 
One consequence from dynamical degeneracy splitting is the uneven broadening in the spectral function at the excitation frequency $\omega$. 
For a given non-Hermitian Hamiltonian $H_{\mathrm{NH}}(\bm{k})$, one can calculate the spectral function $A(\omega,\bm{k})=-\Im\mathrm{Tr}[1/(\omega+i\eta-H_{\mathrm{NH}}(\bm{k}))]$ to characterize the dynamical degeneracy splitting, which can be measured directly, for example, by the Angle-resolved photoemission spectroscopy. 
Therefore, one can identify the dynamical degeneracy splitting from the experimental side by observing the nonuniform broadening of equal-frequency contour under a given excitation frequency. 
Applying this result to condensed matter physics, we further propose that quasiparticle interference become anomalous as discussed in~\cite{SupMat}. 
This will be another experimental signature for the existence of dynamical degeneracy splitting. 

Next, we will demonstrate that anomalous scattering behavior is another consequence of dynamical degeneracy splitting via Green's function. 
The anomalous scattering here refers to the phenomenon that a defect can scatter the propagating plane waves to exponentially damped waves away from the scatterer. 

\begin{figure}[b]
	\begin{center}
		\includegraphics[width=1\linewidth]{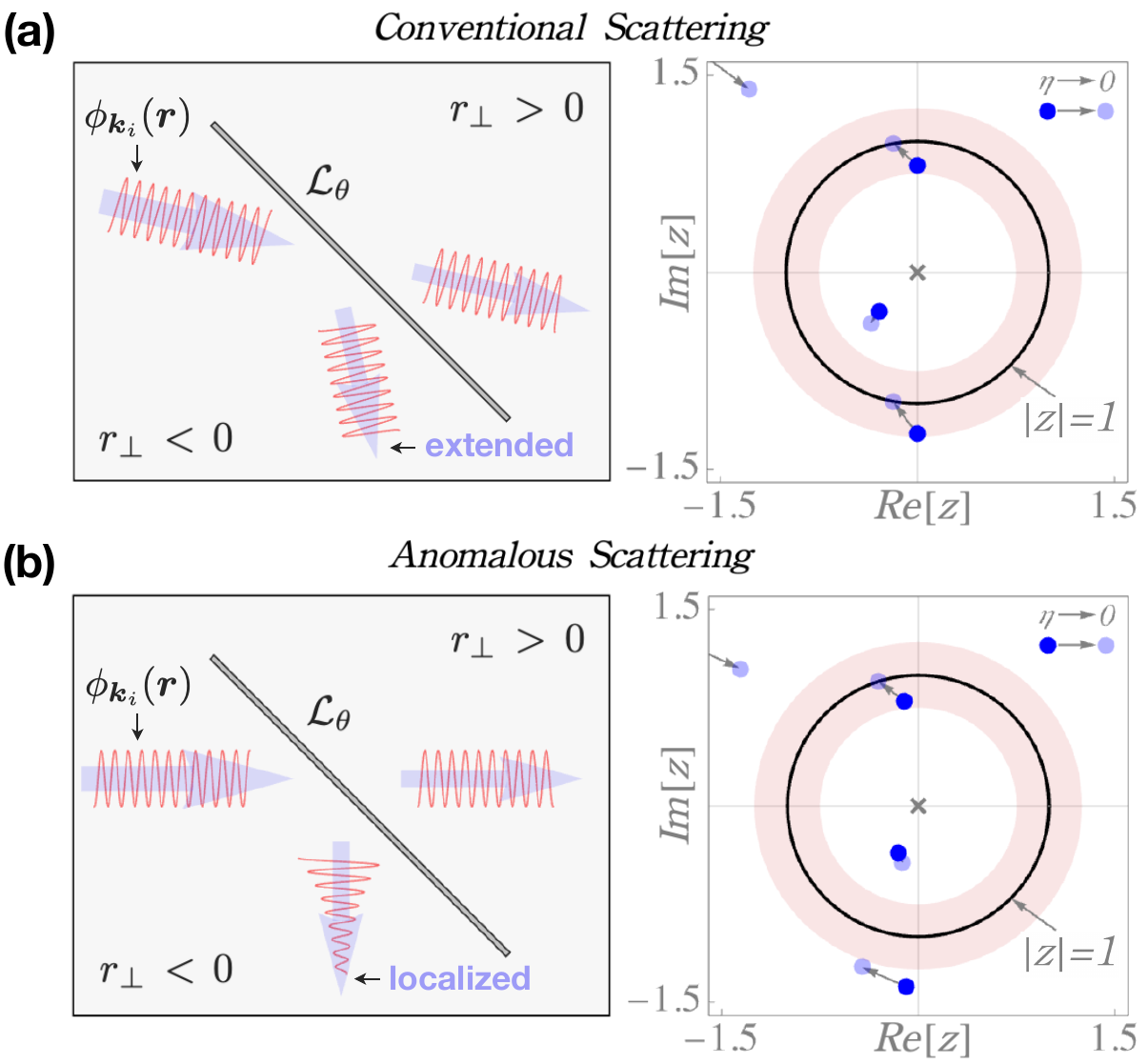}
		\par\end{center}\protect\caption{\label{fig:2} 
		The illustrations of conventional scattering in (a) and anomalous scattering in (b).
		Here, the model Hamiltonian reads $H_{\mathrm{NH}}(\bm{k}) = 2 \sin{k_x} \cos{k_y} - 2 \cos{k_x} + i (\cos{k_x} - 1)$, and the impurity line $\mathcal{L}_{\theta}$ is along $\theta=3\pi/4$ direction. 
		The dark and light blue dots represents parts of poles calculated in Eq.(\ref{MT_ScatWave}). 
		As $\eta\rightarrow 0$, the corresponding poles evolve from dark to light blue dots, as indicated by arrows. } 
\end{figure}

\begin{figure*}[t]
	\begin{center}
		\includegraphics[width=1\linewidth]{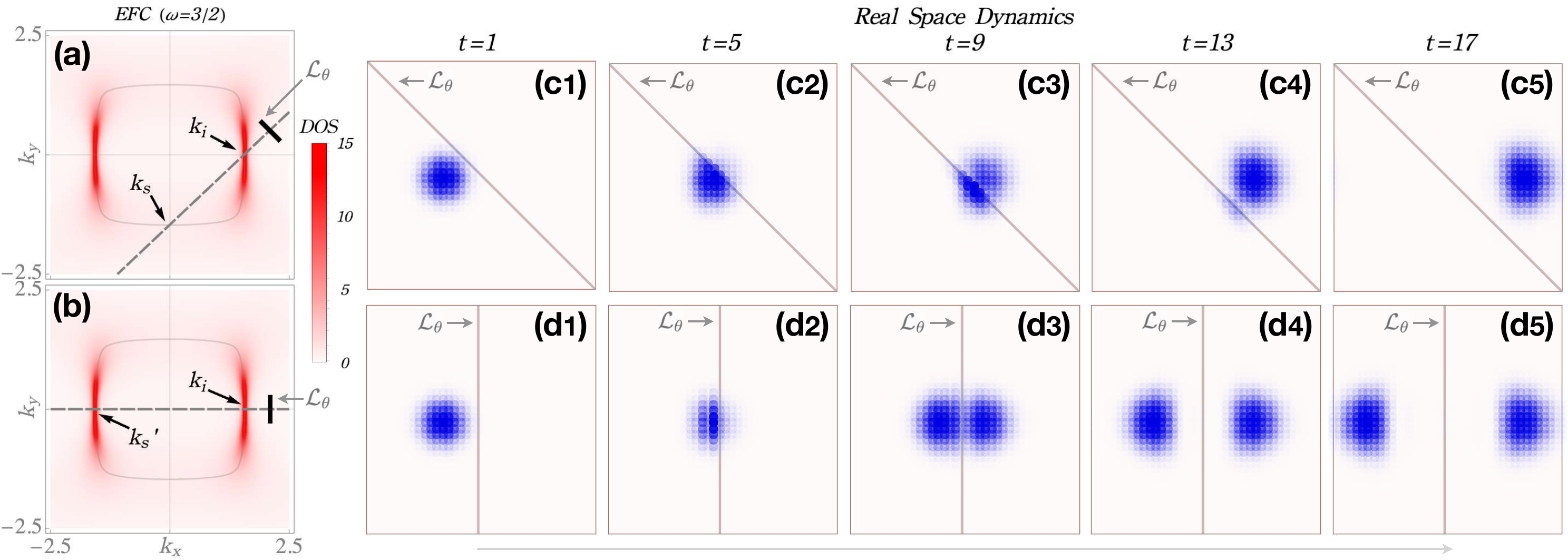}
		\par\end{center}
	\protect\caption{\label{fig:3} 
		Directional invisibility in Hamiltonian Eq.(\ref{MT_ExampleHam}) with the parameters $(\mu_0,\mu_z, t_0, t, t_z, \gamma) = (1.35, -0.05, -0.4, 0.4, -0.6, 1)$. 
		(a-b) show the spectral function $A(\omega,\bm{k})$ with $\omega=3/2$, of which the intensity corresponds to the opacity as shown in the color bar. 
		Here, $\mathcal{L}_{\theta}$ represents the impurity line, and $\bm{k}_i$ denote the incident wave and $\bm{k}_{s}$ ($\bm{k}_{s}^{\prime}$) indicates the scattered wave. 
		The incident wave packet with the momentum center at $\bm{k}_i$ hits the oblique impurity line in (c1)-(c5) and vertical impurity line in (d1)-(d5), where the impurity strength $\lambda=1$. }
\end{figure*}

\emph{{\clr Anomalous scattering theory}.---}~We first establish a general scattering theory in a two-dimensional non-Hermitian system with single band.
It is straightforward to extend our discussion to general situations~\cite{SupMat}. 
The full Hamiltonian can be expressed as, $H=H_{{\mathrm{NH}}}(\bm{k})+V$, where $V$ is the scattering potential. 
Now consider an incident wave $\phi_i(\bm{r})$ (or an excitation) with momentum $\bm{k}_{i}$ propagating on the  lattice and hits the scatterer. 
Then, the scattered waves $\phi_s(\bm{r})$ can be captured by the following integral equation~\cite{SupMat}, 
\begin{equation}\label{MT_InteEqua}
	\begin{split}
		\psi(\bm{r}) &= \phi_{i}(\bm{r}) + \phi_{s} (\bm{r}) \\
		&=\phi_{\bm{k}_i}(\bm{r}) + \int d\bm{r^{\prime}} G_0^{+} (E(\bm{k}_i);\bm{r},\bm{r}^{\prime})\mathcal{V}(\bm{r}^{\prime})\psi(\bm{r}^{\prime}),
	\end{split}
\end{equation}
where $\phi_{\bm{k}_i}(\bm{r}) =\langle \bm{r}|\phi_{\bm{k}_i}\rangle = e^{i\bm{k}_i\bm{r}}$ represents the incident wave, and $\phi_s(\bm{r})$ comprises reflected and transmitted waves. 
Here, $\mathcal{V}(\bm{r})=\langle \bm{r}|V|\bm{r}\rangle$ is the scattering potential function, and $G_0^{+}(E(\bm{k}_i);\bm{r},\bm{r}')=\langle \bm{r}|[E(\bm{k}_i)+i\eta - {H}_{\mathrm{NH}}(\bm{k})]^{-1}|\bm{r}'\rangle$ with $\eta\rightarrow 0^+$ is the retarded Green's function. 
The integral equation Eq.~(\ref{MT_InteEqua}) tells us that: (i) after introducing the scattering potential ${V}$, the eigenstate of the full Hamiltonian ${H}$, i.e. $\psi(\bm{r})$, can be decomposed into the incident and scattered waves with the same energy $E(\bm{k}_i)$~\cite{SupMat}; (ii) the anomalous behavior of the scattering process comes from the anomalous property of the retarded Green's function in non-Hermitian systems. 

Now we use an impurity line, labeled by  $\mathcal{L}_{\theta}$, as an example to demonstrate the anomalous scattering. 
As shown in Fig.~\ref{fig:2}(a)(b), we specify the impurity line lying on the position $r_{\bot}=0$ and along the $\theta$-direction. For the left (right) side of the impurity line $\mathcal{L}_\theta$, the corresponding region is denoted by $r_{\bot}<0$ ($r_{\bot}>0$). 
Therefore, the impurity-line scattering potential function reads
\begin{equation}\label{MT_ImpLineScat}
	\mathcal{V}(\bm{r}) = \lambda \delta(r_\bot=0).
\end{equation} 
Note that the translation symmetry along $r_{\theta}$ direction is preserved in the scattering process. 
Therefore, we substitute Eq.(\ref{MT_ImpLineScat}) into Eq.(\ref{MT_InteEqua}) and take the Fourier transform from $r_{\theta}$ to $k_{\theta}$, and finally obtain the solution of scattering wave~\cite{SupMat} as
\begin{equation}\label{MT_ScatWave}
	\phi_s({k_{i}^{\theta}},r_{\bot}) = \lambda \psi({k_{i}^{\theta}},0) 
	\left\{  
	\begin{split}
		&\sum_{|z_{\mathrm{in}}|<1} {\mathrm{C}}(z_{\mathrm{in}}) z_{\mathrm{in}}^{r_{\bot}}, \,\,\,\,\,\,\,\,\, r_{\bot} > 0; \\  
		&\sum_{|z_{\mathrm{out}}|>1} - {\mathrm{C}}(z_{\mathrm{out}}) z_{\mathrm{out}}^{r_{\bot}}, \,\,\, r_{\bot} < 0,
	\end{split}  
	\right.  
\end{equation}
where $k_{i}^{\theta}=\bm{k}_i\cdot \bm{e}_\theta$ is the $\theta$-component of the incident momentum; the coefficient  $\psi(k_{i}^{\theta},0)$ is a constant for a given $k_{i}^{\theta}$; $\mathrm{C}(z_{\mathrm{in}/\mathrm{out}})$ equals $2\pi i$ times the residue of the function  $[z[E(\bm{k}_i)+i\eta-H_{\mathrm{NH}}(k_i^\theta,z)]]^{-1}$ at the pole $z_{\mathrm{in}/\mathrm{out}}$ inside/outside the $|z|=1$ curve, as shown in Fig.~\ref{fig:2}(a)(b).  

Now we show that when the dynamical degeneracy splitting occurs, the reflected wave will become localized. 
Without loss of generality, we assume the incident plane wave $\bm{k}_i=(k_i^{\theta},k_i^{\bot})$ with energy $E(\bm{k}_i)$ comes from $r_{\bot}<0$ region. 
There are two cases of scattered waves. 
In case (i), there are at least two poles that touch the $|z|=1$ simultaneously from the inner and outer sides respectively when $\eta\rightarrow 0^+$, and one example is illustrated in Fig.~\ref{fig:2}~(a); 
In case (ii), there is only one pole touches the $|z|=1$ curve from the inside as $\eta\rightarrow 0^+$, as shown in Fig.~\ref{fig:2}~(b). 
The more details are present in~\cite{SupMat}.  
It can be seen from Eq.~(\ref{MT_ScatWave}) that there will be two dominant propagating modes survive at infinity for case (i), namely the transmitted wave in the region $r_{\bot}>0$ and the reflected wave in the region $r_{\bot}<0$, as illustrated in Fig.~\ref{fig:2}~(a). 
In case (ii), we have one dominant transmitted wave in the $r_{\bot}>0$ region, while in the $r_{\bot}<0$ region the dominant reflected wave is a spatially localized wave as $\eta \rightarrow 0^+$. 
Therefore, the scattering process is anomalous, as illustrated in Fig.~\ref{fig:2}~(b). 

\emph{{\clr Directional invisibility}.---}~Now we show that for the geometry-dependent skin effect (GDSE), the corresponding scattering process exhibits directional invisibility. 
The Bloch Hamiltonian of GDSE model~\cite{Kai2022NC} reads
\begin{equation}\label{MT_ExampleHam}
	H(\bm{k}) = \sum\limits_{i=0,x,y,z} d_i(\bm{k}) \sigma_i - \frac{i \gamma}{2} (\sigma_0 - \sigma_z),
\end{equation}
where $d_i(\bm{k})$ is real function of $\bm{k}$ and $\sigma_i$ represents the Pauli matrix. 
The only non-Hermitian parameter $\gamma > 0$ is used to describe the dissipative system. 
Specifically, $\{d_0,d_x,d_y,d_z\} = \{  \mu_0 + t_0(\cos {k_x}+\cos {k_y}), t [1-\cos{k_x}-\cos{k_y}+\cos(k_x-k_y)], t [\sin{k_x}-\sin{k_y} - \sin(k_x-k_y)] , \mu_z + t_z (\cos{k_x} - \cos{k_y})  \}$. 
We plot the spectral function $A(\omega_0,\bm{k})$ in Fig.~\ref{fig:3}~(a)(b).
It shows the uneven broadening on the equal-frequency contour (the gray curve representing equal-frequency contour), which is a definite signature of the occurrence of dynamical degeneracy splitting.  
According to the established scattering theory, the anomalous scattering will occur. 

We assume an incident plane wave has $\bm{k}_i=(k_i^{x},k_i^{y})=(\pi/2,0)$ and hits the impurity line $\mathcal{L}_{\theta}$ with a rightward velocity in real space. 
Note that $\bm{k}_i$ lies on the equal-frequency contour, i.e., $\bm{K}(\omega)$ with $\omega =3/2$, as shown in Fig.~\ref{fig:3}(a)(b). 
The impurity line $\mathcal{L}_{\theta}$ preserves the momentum along this direction, which means that the scatterer $\mathcal{L}_{\theta}$ relates $\bm{k}_i$ with $\bm{k}_{s}$ and $\bm{k}_{s}^{\prime}$ in the way illustrated in Fig.~\ref{fig:3}(a) and (b), respectively. 
In Fig.~\ref{fig:3}(a), due to the larger broadening at $\bm{k}_{s}$ than that at $\bm{k}_{i}$, the reflected wave is damped exponentially away from the impurity line, which means that anomalous scattering occurs as discussed  in case (ii). 
The band dispersion of the Hamiltonian in Eq.(\ref{MT_ExampleHam}) is mirror symmetric under $\mathcal{M}_x$: $(k_x,k_y)\rightarrow (-k_x,k_y)$ and $\mathcal{M}_y$: $(k_x,k_y)\rightarrow (k_x,-k_y)$. 
This means $\bm{k}_{s}^{\prime}=\mathcal{M}_x \bm{k}_i$ has the equal broadening with $\bm{k}_i$, as shown in Fig.~\ref{fig:3}(b). 
Therefore, the conventional scattering discussed in the case (i) occurs for the vertical impurity line that scatters the incident plane wave to another propagating plane wave $\bm{k}_{s}^{\prime}$. 
This phenomenon that the visibility of reflected waves depends on the direction of impurity line is dubbed {\em directional invisibility} and unique to higher-dimensional non-Hermitian systems. 

To probe the directional invisibility in this example, the incident wave is chosen as a Gaussian wave packet with momentum center at $\bm{k}_i$ 
for the scattering simulation, as shown in Fig.~\ref{fig:3}(c)(d). 
The time evolution of wave packet follows $|\psi(t)\rangle = \mathcal{N}(t)e^{- i H t} |\phi_0\rangle$, where $H$ is the full Hamiltonian consisting of the free Hamiltonian in Eq.(\ref{MT_ExampleHam}) and impurity-line scattering potential $\mathcal{V}(\bm{r})=\lambda \sigma_0 \delta(r_{\bot}=0)$, and $\mathcal{N}(t)$ is the normalization factor at every time.  
The incident Gaussian wave packet has the form $\phi_0(\bm{r}) = \mathrm{exp}[-(\bm{r}-\bm{r}_0)^2/\sigma^2+i \bm{k}_i^0 \bm{r}](1,1)^T$. 
In Fig.~\ref{fig:3}(c)(d), the parameters are set as $(x_0,y_0,\sigma)=(14,20,4)$, and the lattice size is $L_x=L_y=40$. 
It can be observed that the Gaussian wave packet is almost completely transmitted through the oblique impurity line $\mathcal{L}_{\theta}$ without evident reflected waves, as shown in Fig.~\ref{fig:3}(c1)-(c5). 
However, parts of the wave packet are reflected by the vertical impurity line $\mathcal{L}_{\theta}$ as a propagating wave, shown in Fig.~\ref{fig:3}(d1)-(d5). 
Therefore, the wave-packet scattering simulation in Fig.~\ref{fig:3}(c)(d) verifies the directional invisibility. 

\emph{{\clr The role of symmetry}.---}~Now we discuss the role of symmetry, which reveal the correspondence between directional invisibility and GDSE. 
For $H_{\mathrm{NH}}(\bm{k})$, all symmetries preserving the complex energy form the scattering group of $H_{\mathrm{NH}}(\bm{k})$, labeled by $G_s$, which includes, for example, the reciprocity $\mathcal{\bar{T}}$~\cite{Sato2019_PRX} and point-group symmetries, such as rotation, inversion, and mirror symmetry $\mathcal{M}$. 
Now suppose that there is an incident wave with momentum $\bm{k}_i$. 	
Under the action of $G_s$, $\bm{k}_i$ will be mapped to a set of other points on the BZ with the same energy, that is, $E_{\mu}(\bm{k}_i) = E_{\mu}(g_s\bm{k}_i)$ with $g_s\in G_s$. 
Note that $\bm{k}_i$ and $g_s\bm{k}_i$ determines a direction crossing them, and we label the impurity line perpendicular to this direction by $\mathcal{L}_{g_s\bm{k}_i}$. 
Now we state the conclusion: 
for the incident wave $\bm{k}_i$, the scattering process for the impurity line $\mathcal{L}_{g_s\bm{k}_i}$ is conventional; while for all other directions, the scattering process is not protected by symmetry $g_s$ and is generally anomalous. 
It should be noted that if $g_s=\mathcal{M}$ or $\bar{\mathcal{T}}\mathcal{M}$, $\mathcal{L}_{g_s\bm{k}_i}$ is exactly parallel or perpendicular to the mirror line, respectively, and is independent of $\bm{k}_i$. 
We label such a impurity line as $\mathcal{L}_{g_s}$. 
It means that for the impurity line $\mathcal{L}_{g_s}$, the scattering process for all possible incident states is conventional. 
For example, if $H_{\mathrm{NH}}(\bm{k})$ preserves $\mathcal{M}_x$ symmetry, then $\mathcal{L}_{\mathcal{M}_x}$ is along the $y$-direction, and the conventional scattering occurs on the impurity line $\mathcal{L}_{\mathcal{M}_x}$ for all possible incident waves. 

Now we relate it to the GDSE. 
For GDSE, if there is an edge parallel to the impurity line $\mathcal{L}_{g_s}$, then the edge shows conventional scattering for all $\bm{k}_i\in \mathrm{BZ}$, correspondingly, the open boundary eigenstates cannot be localized at that edge. 
Based on this principle, one can find that if the Hamiltonian has $\mathcal{M}$ and/or $\bar{\mathcal{T}}\mathcal{M}$ symmetry, then open boundary eigenstates can no longer be localized at the edges parallel to $\mathcal{L}_{\mathcal{M}}$ and/or $\mathcal{L}_{\bar{\mathcal{T}}\mathcal{M}}$ under any shape of open boundary geometry. 

\emph{{\clr Conclusions and Discussions}.---}~In summary, we introduce the concept of dynamical degeneracy splitting to characterize the nonuniform decay behavior of excited modes at a given frequency. 
On the one hand, the dynamical degeneracy splitting predicts the frequency-resolved NHSE; on the other hand, it associates with the anomaly in Green's function at a specified frequency, leading to uneven broadening in spectral function and anomalous scattering. 
As an application, we propose a type of anomalous scattering, directional invisibility, as an experimental indicator of the existence of GDSE.

This work essentially provides a frequency criterion that can further help us understand and define NHSE in the system beyond conventional band theory. 
For example, in the electronic system with self-energy corrections, the retarded Green function has the form $G(\omega,\bm{k}) = [\omega - \mathcal{H}_{\mathrm{eff}}(\omega,\bm{k})]^{-1}$, where the effective Hamiltonian depends on the frequency $\omega$. 
The dynamical degeneracy splitting can still be well-defined, correspondingly, the concept of NHSE can be extended in such systems, which laid the foundation for further studies. 

\let\oldaddcontentsline\addcontentsline
\renewcommand{\addcontentsline}[3]{}
\input{DDSDI.bbl}
\let\addcontentsline\oldaddcontentsline
\onecolumngrid
\newpage
\makeatletter
\renewcommand \thesection{S-\@arabic\c@section}
\renewcommand\thetable{S\@arabic\c@table}
\renewcommand \thefigure{S\@arabic\c@figure}
\renewcommand \theequation{S\@arabic\c@equation}
\makeatother
\setcounter{equation}{0}  
\setcounter{figure}{0}  
\setcounter{section}{0}  
{\begin{center}
		{\bf \large Supplemental Material for ``Dynamical Degeneracy Splitting and Directional Invisibility in Non-Hermitian Systems'' }
\end{center}}
\maketitle
\tableofcontents

\section{Equal frequency contour and dynamical degeneracy splitting}

In this section, we will give a general method for obtaining the equal-frequency contour in non-Hermitian systems. 
As an example, we use a two-band model to demonstrate the equal-frequency contour and dynamical degeneracy splitting.  
Then, we discuss the cases where singularities and intersections exist on the equal-frequency contour. 
In the last part, we give the relation between dynamical degeneracy splitting and the emergence of non-Hermitian skin effect. 

\subsection{General formula of equal frequency contour in non-Hermitian systems}

We assume a (non-Hermitian) Bloch Hamiltonian $\mathcal{H}(\bm{k})$ with $m$ bands in $d$ dimensions, of which the characteristic equation can be expressed as 
\begin{equation}\label{S1A_CharacEqua}
	f(\omega, \mathrm{Im}E,\bm{k})=\det[\mathcal{H}(\bm{k}) - (\omega+i\mathrm{Im}E)\, \mathbb{I}_{m}]=0,
\end{equation}
where $E=\omega+i \mathrm{Im}E$ is the complex energy with the excitation frequency $\omega$ and line width $\mathrm{Im}E$, and $\omega,\mathrm{Im}E \in \mathbb{R}$. 
$\mathbb{I}_{m}$ represents the $m\times m$ identity matrix. 
In general, $f(\omega,\mathrm{Im}E,\bm{k})$ is a complex function, hence,  Eq.(\ref{S1A_CharacEqua}) imposes two real constraints, $f_r(\omega,\mathrm{Im}E,\bm{k})=f_i(\omega,\mathrm{Im}E,\bm{k})=0$.
Here, the subscript $r/i$ represents the real/imaginary part of $f(\omega,\mathrm{Im}E,\bm{k})$, respectively. 
By definition, equal-frequency contour is only determined by $\omega$ and is regardless of $\mathrm{Im}E$. 
Therefore, we can give an analytic traceable formula to equal-frequency contour. 
We define
\begin{equation}\label{S1A_ResulantEFC}
	F(\omega,\bm{k}) := \mathrm{Res}[f_r(\omega,\mathrm{Im}E,\bm{k}), f_i(\omega,\mathrm{Im}E,\bm{k}), \mathrm{Im}E ]. 
\end{equation}
Here, $\mathrm{Res}[f_r, f_i, \mathrm{Im}E]$ represents resultant~\cite{SM_Zhesen2020} between $f_r$ and $f_i$ regarding variable $\mathrm{Im}E$. 
After doing this, variable $\mathrm{Im}E$ is eliminated and a real function $F(\omega,\bm{k})$ is obtained. 
Finally, for a generic $\omega_0$, we can obtain the ($d-1$)-dimensional equal-frequency contour expressed as, 
\begin{equation}\label{S1A_GenericEFC}
	\bm{K}(\omega_0)=\{\bm{k}\in {\rm BZ} |  F(\omega_0,\bm{k})=0  \}.
\end{equation}
Here the expression of equal-frequency contour hides the information of the energy band index, which is slightly different from the definition in the main text. 
Note that for each $\bm{k}\in \mathrm{BZ}$, the energy band index is well defined.
Therefore, we can obtain the expression of equal-frequency contour in the main text, 
\begin{equation}\label{MT_EFCDef}
	\bm{K}(\omega\in\mathbb{R})=\bm{K}_1(\omega\in\mathbb{R})\cup...\cup\bm{K}_m(\omega\in\mathbb{R}),
\end{equation}
where $\bm{K}_\mu(\omega\in\mathbb{R})=\{\bm{k}\in {\rm BZ} | \Re E_\mu(\bm{k})=\omega\}$, and $E_{\mu}(\bm{k})$ represents the $\mu$-th band of the Bloch Hamiltonian $\mathcal{H}(\bm{k})$. 

Here, we discuss the physical meaning of equal-frequency contour with dynamical degeneracy splitting. 
As a general assumption, equal-frequency contour is always formed by simple closed curves without singularities or intersections. 
It means that $\nabla_{\bm{k}}\Re E_\mu(\bm{k})|_{\bm{k}\in \bm{K}_\mu(\omega)}\neq 0$ for $\mu=1,...,m$, which is satisfied for a generic frequency $\omega$. 
For a given frequency $\omega_0$, each point on the equal-frequency contour corresponds to an excited mode, and the group velocity in real space is along the normal direction at that point, which can be calculated by
\begin{equation}\label{SM_GroupVelocity}
	\bm{v}(\bm{k}) = (v_{x},v_{y}) = \left. \left( -\frac{\partial_{k_x} F(\omega,\bm{k})}{\partial_{\omega} F(\omega,\bm{k})} ,  -\frac{\partial_{k_y} F(\omega,\bm{k})}{\partial_{\omega} F(\omega,\bm{k})}\right) \right|_{\omega=\omega_0},
\end{equation}
where $F(\omega,\bm{k})$ in defined in Eq.(\ref{S1A_ResulantEFC}). 
When dynamical degeneracy splitting occurs, the excited modes on $\bm{K}_{\mu}(\omega)$ obtains different decay rates and will evolve as $\sim e^{-\mathrm{Im}E_{\mu}(\bm{k})t}$. From the perspective of spectral information, the imaginary part $\Im E_\mu(\bm{k})$ will broaden the equal-frequency contour. 
This can be usually reflected in spectral function, 
\begin{equation}\label{SM_SpecFunc}
	A(\omega,\bm{k}) = - \mathrm{Im} \mathrm{Tr} [(\omega + i \eta - \mathcal{H}(\bm{k}) )^{-1}]/N,
\end{equation}
where $\eta\rightarrow 0^+$ and $N$ is the total number of energy bands, therefore, it is detectable in experiments.  

\begin{figure}[t]
	\begin{center}
		\includegraphics[width=1\linewidth]{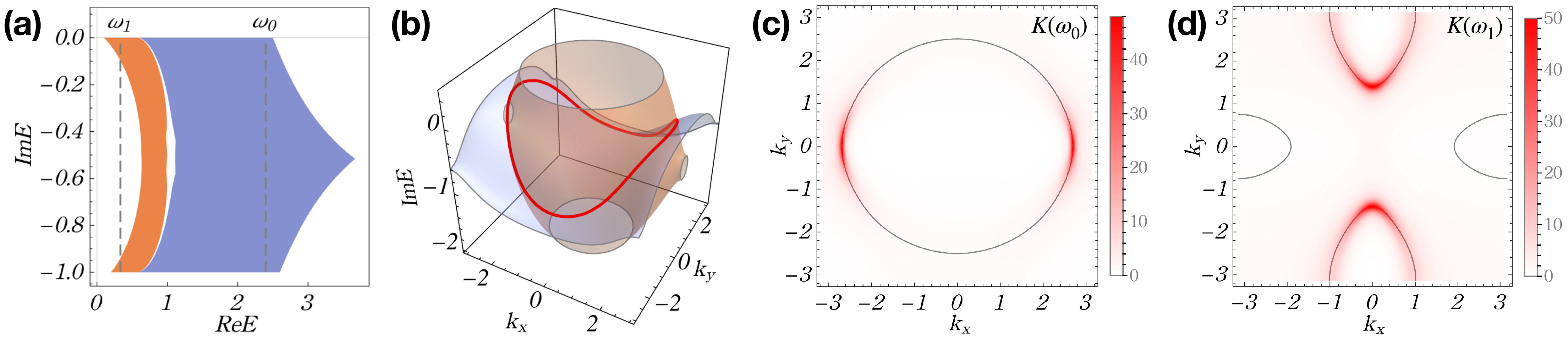}
		\par\end{center}
	\protect\caption{\label{SM_FigS1} Non-Hermitian Bloch spectrum (a), equal-frequency contour and dynamical degeneracy splitting (b-d) for the Hamiltonian Eq.(\ref{SM_TBModel}) with system parameters $\{ \mu_0,\mu_z, t_0, t, t_z, \gamma \} = \{ 1.35, -0.05, -0.4, 0.4, -0.6, 1 \}$. The red and blue surface in (b) represent $f_r(\omega_1,\mathrm{Im}E, \bm{k})=0$ and $f_i(\omega_1,\mathrm{Im}E, \bm{k})=0$, respectively. The gray solid lines in (c) and (d) represent the equal-frequency contours at $\omega_0$ and $\omega_1$, respectively. The spectral function in Eq.(\ref{SM_SpecFunc}) is plotted in (b)(c) with $\eta=1/50$, and the intensity of spectral function corresponds to the opacity as shown in the color bar. } 
\end{figure}

\subsection{An example in two-band model}

As an example, we consider the two-band Bloch Hamiltonian in Eq.(5) in the main text, that is, 
\begin{equation}\label{SM_TBModel}
	\mathcal{H}(\bm{k}) = \sum_{i=0,x,y,z} d_i(\bm{k}) \sigma_i - \frac{i \gamma}{2} (\sigma_0 - \sigma_z),
\end{equation}
where $\{d_0,d_x,d_y,d_z\} = \{  \mu_0 + t_0(\cos {k_x}+\cos {k_y}), t [1-\cos{k_x}-\cos{k_y}+\cos(k_x-k_y)], t [\sin{k_x}-\sin{k_y} - \sin(k_x-k_y)] , \mu_z + t_z (\cos{k_x} - \cos{k_y})  \}$. 
This model demonstrates the bulk Fermi arc~\cite{SM_Zhou2018} in photonic crystal experiment. 
Theoretically, this model has the geometry-dependent skin effect~\cite{SM_Kai2022NC}, i.e., non-Hermitian skin effect appears under open boundary conditions of generic geometries (e.g., the diamond geometry), but disappears under square geometry. 

The Bloch spectrum of Hamiltonian Eq.(\ref{SM_TBModel}) is shown in Fig.~\ref{SM_FigS1}(a). 
We choose $\omega_0$ and $\omega_1$ as the excitation frequency, as indicated by the dashed lines. 
When $\omega=\omega_0$, we can get two surfaces in $(k_x,k_y,\mathrm{Im}E)$ space, namely $f_r(\omega_1, \mathrm{Im}E, \bm{k})=0$ and $f_i (\omega_1, \mathrm{Im}E, \bm{k})=0$, as shown by the red and blue surfaces in Fig.~\ref{SM_FigS1}(b), respectively. 
The intersection of these two surfaces is shown by the red line in Fig.~\ref{SM_FigS1}(b), and its projection onto $\bm{k}$ plane is exactly the equal-frequency contour curve in (c). 
Actually, according to the formula in Eq.(\ref{S1A_GenericEFC}), we can directly obtain the equal-frequency contours $\bm{K}(\omega_0)$ and $\bm{K}(\omega_1)$, as plotted by the gray solid lines in Fig.~\ref{SM_FigS1}(c) and (d), respectively. 
It can be seen from Fig.~\ref{SM_FigS1}(b) that the momenta on the equal-frequency contour have different imaginary energies $\mathrm{Im}E(\bm{k})$, which means that dynamical degeneracy splitting occurs. 
It can be reflected in the spectral function $A(\omega,\bm{k})$. 
We plot the spectral function $A(\omega_0,\bm{k})$ and $A(\omega_1,\bm{k})$ in Fig.~\ref{SM_FigS1}(c) and Fig.~\ref{SM_FigS1}(d), respectively.
The intensity of spectral function corresponds to the opacity of red color.
It demonstrates the occurrence of dynamical degeneracy splitting in this two-band model. 

\subsection{The singularities and intersections on equal-frequency contour}

In this part, we first demonstrate that equal-frequency contour is always formed by simple closed curves without singularities or intersections.
Then, we discuss the case where exceptional points exist on the equal-frequency contour and give an example to show it.  

As we defined in Eq.(\ref{S1A_ResulantEFC}), the algebraic equation that determines the equal-frequency contour is 
\begin{equation*}
	F(\omega, \boldsymbol{k}):=\operatorname{Res}\left[f_r(\omega, \operatorname{Im} E, \boldsymbol{k}), f_i(\omega, \operatorname{Im} E, \boldsymbol{k}), \operatorname{Im} E\right]=0, 
\end{equation*}
which is a real function with variables $\bm{k}$ and $\omega$. Mathematically, singularities or intersections on the equal-frequency contour must satisfy
\begin{equation}\label{EFC1}
	F(\omega,\bm{k})=\nabla_{\bm{k}}F(\omega,\bm{k})=0. 
\end{equation}
In 2D cases, Eq.(\ref{EFC1}) reduces to $F(\omega,\bm{k})=\partial_{k_x}F(\omega,\bm{k})=\partial_{k_y}F(\omega,\bm{k})=0$. 
Since we have three real variables $\omega, k_x, k_y$, and three real equations, the dimension of corresponding solutions is zero-dimension, which is nothing but a set of points. The $\omega$ in the solution determines the corresponding equal-frequency contour with singularities or intersections. Since the solution is a set of points, for a generic excitation frequency $\omega_0$, it is expected that it does not include singularities or intersections. 
Therefore, we assume that equal-frequency contour is always formed by simple closed curves without singularities or intersections. 

In general, the non-Hermitian band degeneracy is determined by the following equations
\begin{equation}\label{EP}
	f(E, \bm{k})=\partial_E f(E, \bm{k})=0. 
\end{equation}
Here $E\in\mathbb{C}$ and $f(E,\bm{k})$ is a complex equation with two real components. In general, the solution of the above equation determines the exceptional points in the Bloch bands. 

For a given equal-frequency contour band that satisfies Eq.(\ref{EFC1}) $F(\omega,\bm{k})=0$, the corresponding band degeneracy is determined by the following equations 
\begin{equation}\label{dEFC}
	F(\omega,\bm{k})=\partial_\omega F(\omega,\bm{k})=0. 
\end{equation} 
For 2D cases, we have three real variables and two real equations.
Therefore, the corresponding solution is general a 1D line in the parameter space. 
This solution exactly determines the Fermi arc in the BZ. 
In general, the end point of the Fermi arc is the exceptional point. 
Therefore, the solution of Eq.(\ref{EP}) in general belongs to Eq.(\ref{dEFC}).
Actually, we can find the common solutions between Eq.(\ref{EFC1}) and Eq.(\ref{dEFC}), which are determined by $F(\omega,\bm{k})=\partial_\omega F(\omega,\bm{k})=\nabla_{\bm{k}}F(\omega,\bm{k})=0$. 
These points are singularities of $F(\omega,\bm{k})$. 

\begin{figure}[t]
	\begin{center}
		\includegraphics[width=1\linewidth]{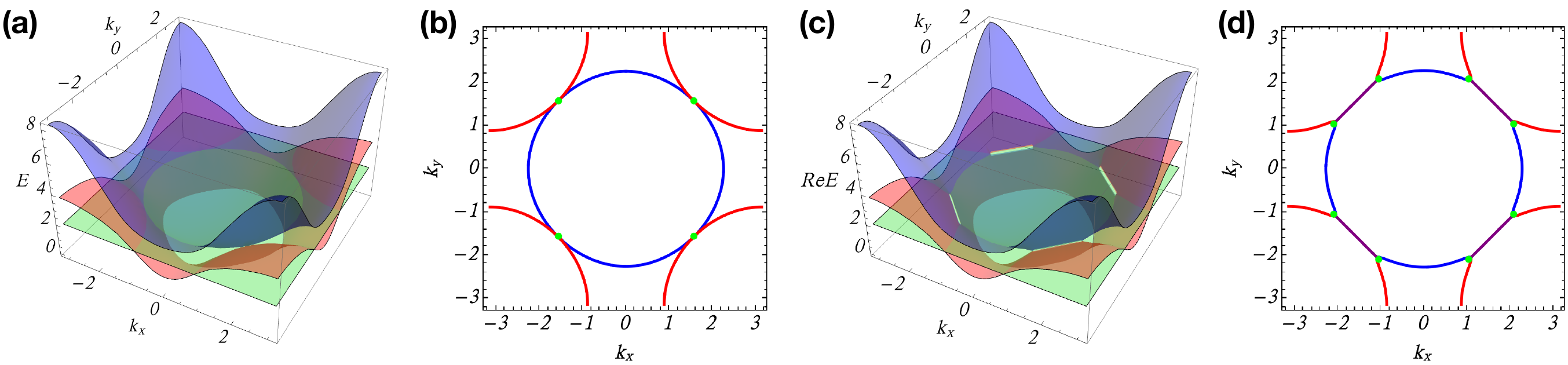}
		\par\end{center}
	\protect\caption{\label{SM_FigS7} (a) The band structure and (b) equal-frequency contour at $\omega=2$ of Hamiltonian Eq.(\ref{R1_EPModel}) when $\gamma=0$. 
		The green plane in (a) refers to $E=0$ plane and the green dots in (b) corresponds to the band degeneracies. 
		(c) The band structure and (d) equal-frequency contour at $\omega=2$ of Hamiltonian Eq.(\ref{R1_EPModel}) when $\gamma=1/4$. 
		The green dots and blend-color arcs in (d) represent exceptional points and bulk Fermi arcs, respectively. 
	}
\end{figure}

Here, we take a two-band example that has equal-frequency contour with exceptional points and Fermi arcs. 
The Hamiltonian of this example reads 
\begin{equation}\label{R1_EPModel}
	\mathcal{H}(\bm{k}) = (2-2\cos{k_x}-2\cos{k_y})\,\sigma_0 
	+ \cos{k_x}\cos{k_y} \, \sigma_x - (\cos{k_x}+\cos{k_y}) \, \sigma_z 
	- i \gamma \, (\sigma_0-\sigma_z),
\end{equation} 
where $\sigma_{x,y}$ represent Pauli matrices and $\sigma_{0}$ is the identity matrix. 
The last term is the non-Hermitian term, meaning that the second orbital has a decay rate $\gamma$. 
First, let $\gamma$ be zero, the Hamiltonian reduces to Hermitian and has the band structure plotted in Fig.~\ref{SM_FigS7}(a), where the blue (red) surface indicates the first (second) energy band and the green plane refers to the energy plane $E=2$. 
The corresponding equal-frequency contour at $\omega_0=2$ is shown in Fig.~\ref{SM_FigS7}(b), where the blue (red) curve corresponds to the band with the same color. 
Note that on the equal-frequency contour at $\omega_0$, there are four degeneracies indicated by green dots in Fig.~\ref{SM_FigS7}(b). 

When adding the loss term ($\gamma=1/4$), the real energy bands of the non-Hermitian Hamiltonian $\mathcal{H}(\bm{k})$ are plotted in Fig.~\ref{SM_FigS7}(c), and the corresponding equal-frequency contour at $\omega_0=2$ is shown in Fig.~\ref{SM_FigS7}(d). 
It shows that the original degeneracies split into eight exceptional points (the green dots) that are connected by four pieces of bulk Fermi arcs (in blend color). 
On the Fermi arcs, two real energy bands degenerate but the imaginary parts are different. 
The group velocity at exceptional points cannot be defined, similar to the intersections. 
Note that in this example, only the equal-frequency contour at $\omega=2$ has exceptional points. 
When we excite at the frequency of Fermi arcs, there are always two modes are excited with velocities,  $\bm{v}_{\mu=1,2}=(\partial_{k_x}\Re{E}_{\mu},\partial_{k_y}\Re{E}_{\mu})$, and these two excited modes must have different decay rates. 

\subsection{The relationship between dynamical degeneracy splitting and non-Hermitian skin effect}

Here, we discuss the relation between dynamical degeneracy splitting and non-Hermitian skin effect by the following logic: 
(i) The appearance of dynamical degeneracy splitting; 
(ii) The original linear superposition of Bloch waves is no longer to be the eigenstate of $H_{\mathrm{OBC}}$;
(iii) If $E_0$ is the eigenvalue of $H_{\mathrm{OBC}}$, 
the corresponding eigenstate must contain some non-Bloch wave components. 
The appearance of non-Bloch waves in the OBC eigenstate implies the emergence of non-Hermitian skin effect;
(iv) If $E_0$ is not the eigenvalue of $H_{\mathrm{OBC}}$, the mismatch between PBC and OBC spectrum also indicates the emergence of non-Hermitian skin effect~\cite{SM_Kai2020,SM_Okuma2020}.
Therefore, no matter whether $E_0$ is the OBC eigenvalue or not, the non-Hermitian skin effect always appears. 

Now we explain it in more details. 
In order to simplify the discussion, we consider a single band model with the Hamiltonian $$H({\bm{k}})=h_0(\bm{k})+i\lambda\Gamma_0(\bm{k}),$$ 
where $h_0(\bm{k})$ and $\Gamma_0(\bm{k})$ are real functions. 
For a given excitation energy $\omega_0\in\mathbb{R}$, the corresponding equal-frequency contour $\bm{K}(\omega_0)$ is defined as $$\bm{K}(\omega_0)=\{\bm{k}\in {\rm{BZ}}|\omega_0=\Re{H}(\bm{k}) \}.$$
And the set of pre-images of energy $E_0$ in BZ is defined as $$\bm{Q}(E_0)=\{\bm{k}\in {\rm{BZ}}|E_0= H(\bm{k})\}.$$
These $\bm{k}$ (plane waves) in the set $\bm{Q}(E_0)$ can be linearly superimposed into the wavefunction with energy $E_0$ under some particular OBC geometry (for example, square geometry), that is, 
\begin{equation}\label{SupMat_BlukWave}
	\psi_0(\bm{r}) = \sum_{\bm{k}\in \bm{Q}(E_0)} c_{\bm{k}} e^{i \bm{k} \bm{r}},
\end{equation}
where these independent coefficients $c_{\bm{k}}$ can be determined by specific open boundary conditions in a 2D lattice. 
Note that the wavefunction $\psi_0(\bm{r})$ compose of plane waves is always extended wavefunction. 
By definition, the equal-frequency contour $\bm{K}(\omega_0\in \mathbb{R})$ is always 1D line in BZ, while $\bm{Q}(E_0)$ is not always 1D line. 

We consider the non-Hermitian case with $\lambda\neq 0$. 
Owing to the term $i\lambda \Gamma_0(\bm{k})$, dynamical degeneracy splitting may occur, and the equal-frequency contour will split in term of imaginary energy. 
Correspondingly, the degenerate eigenvalue $\omega_0$ in Hermitian limit splits into $\omega_0+i\lambda\Gamma_0(\bm{k})$ with $\bm{k} \in\bm{K}(\omega_0)$. 
Taking one of the eigenvalues as an example, $E_0=\omega_0+i\gamma_0$.
The corresponding pre-images in BZ $\bm{Q}(E_0)$ include some (order-1) $\bm{k}$ points. 
However, for a two-dimensional lattice, there are order-$L$ different open boundary conditions ($L$ represents the system length); for example,  there are order-$L$ edges in different angles. 
Therefore, we need order-$L$ independent coefficients $c_{\bm{k}}$ such that the extended wavefunction $\psi_0(\bm{r}) $ with energy $E_0=\omega_0+i\gamma_0$ can always be superimposed under all order-$L$ possible open boundary conditions. 
Otherwise, there must be some open boundary (geometries) conditions under which the wave function $\psi_0(\bm{r})$ with $E_0$ is a localized skin mode instead of an extended wave. 
Finally, we conclude that $\bm{Q}(\omega_0+i\gamma_0)$ having some (order-1) $\bm{k}$ points is insufficient to form extended wavefunction $\psi_0(\bm{r})$ in Eq.(\ref{SupMat_BlukWave}) under all order-$L$ different open boundary (geometries) conditions. 
Therefore, for a wave function $\psi_0$ with energy $E_0=\omega_0+ i \gamma$ under generic open boundary geometry, if we assume it to be an extended wavefunction, the number of plane waves $\bm{k}$ to form the extended wavefunction will in general be order-L in two dimensions. 
In the case of dynamical degeneracy splitting, the set $\bm{Q}(\omega_0+i\gamma_0)$ is not sufficient to form the Bloch wave under a generic OBC geometry. 
It implies that if $\omega_0+i\gamma_0$ is the OBC eigenvalue, the corresponding OBC eigenstate must contain some non-Bloch components with $\Im \bm{k}\neq 0$, which means the appearance of non-Hermitian skin effect. 
If $\omega_0+i\gamma_0$ is not the OBC eigenvalue, the mismatch between PBC and OBC spectrum also indicates the emergence of non-Hermitian skin effect. 
Therefore, no matter whether $E_0=\omega_0+i\gamma_0$ is the OBC eigenvalue or nor, the non-Hermitian skin effect always appears. 

\subsection{Quasiparticle interference and dynamical degeneracy splitting}

\begin{figure}[b]
	\begin{center}
		\includegraphics[width=.8\linewidth]{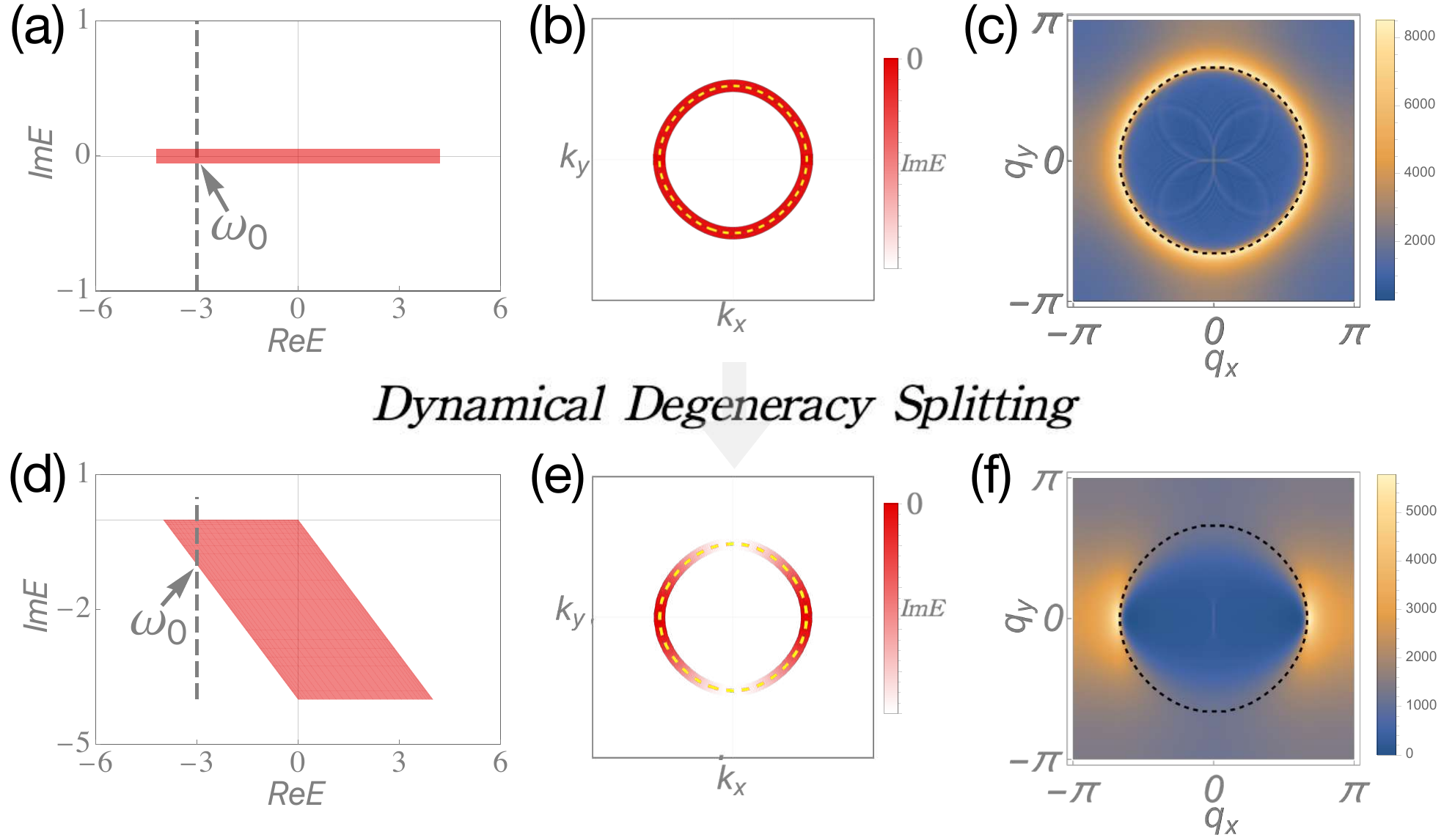}
		\par\end{center}
	\protect\caption{\label{Fig1RC3} 
		(a)-(c) and (d)-(f) show the spectrum, equal-frequency contour weighted by different imaginary energies, and quasiparticle interference pattern for the Hamiltonian in Eq.(\ref{RB_Model}) with $\gamma=0$ and $\gamma=1$, respectively. 
		In (b)(e) the yellow dashed circle represents equal-frequency contour at $\omega_0=-3$, and the color bar corresponds to the $\Im E(\bm{k})$ on the equal-frequency contour. 
		In (c)(f) the color bar corresponds to the quantity $\rho(\bm{q},\omega_0)$ calculated in Eq.(\ref{RB_QPI})
		.} 
\end{figure}

In condensed matter physics, the interference pattern of quasiparticles around a single impurity observed through scanning tunneling spectroscopy has been widely used for electronic structure characterization of unconventional states, for example, the topological surface states. 
Consider a point impurity in the system, the local density of states is no longer uniform due to the breakdown of translation symmetry. 
The quasiparticle interference pattern can be calculated by the Fourier transform of the local density of states. 
It can be derived as 
\begin{equation}\label{RB_QPI}
	\rho(\bm{q},\omega) = \frac{1}{N \pi} \sum_{\bm{k}} \mathrm{Tr}[G(\bm{k},\bm{k+q},\omega)-G^{\ast}(\bm{k+q},\bm{k},\omega)], 
\end{equation}
where $G(\bm{k},\bm{k^{\prime}},\omega)$ represents the retarded Green's function including the impurity potential. 
The position of high-intensity peaks in $\rho(\bm{q},\omega)$ generically depends on the geometric features of the equal-frequency contour at frequency $\omega$. 
As we pointed out in our manuscript, the dynamical degeneracy splitting leads to the anomalous scattering behaviors. 
Therefore, we can expect dynamical degeneracy splitting will significantly affect the quasiparticle interference pattern, which can be directly detected in scanning tunneling spectroscopy experiments. 

Next, we use a simple example to demonstrate it. 
The Hamiltonian of the simple model can be written as 
\begin{equation}\label{RB_Model}
	H_{\mathrm{NH}}(\bm{k})=-2 \cos{k_x} - 2 \cos{k_y} - 2 i \gamma (1 - \cos{k_y}),
\end{equation}
where $\gamma$ controls the non-Hermitian part that leads to the dynamical degeneracy splitting.  
We can clearly observe in Fig.~\ref{Fig1RC3}(a)(b) that when $\gamma=0$, there is no dynamical degeneracy splitting at $\omega_0=-3$, and consequently, the highest intensity peaks in the quasiparticle interference pattern in Fig.~\ref{Fig1RC3}(c) reveal the dominant scattering channels on the equal-frequency contour at $\omega_0$, that is, $\bm{q}=2\bm{k}$ corresponding to the back scattering in this simple example. 
The secondary intensity peaks looks like petal shape inside the dashed black circle in Fig.~\ref{Fig1RC3}(c) and corresponds to the scattering channels from $\bm{k}$ to $\bm{k} e^{i \theta}$ ($\theta \neq 0, \pi$) on the equal-frequency contour in Fig.~\ref{Fig1RC3}(b) . 

As shown in Fig.~\ref{Fig1RC3}(d)(e), when $\gamma = 1$, dynamical degeneracy splitting occurs at $\omega_0=-3$. 
Consequently, the quasiparticle interference pattern in Fig.~\ref{Fig1RC3}(f) significantly different from that in Fig.~\ref{Fig1RC3}(c). 
We conclude the following main points from the differences in their quasiparticle interference patterns. 
First, the highest peaks Fig.~\ref{Fig1RC3}(f) reveal that the dominant scattering channels occur between the momenta with longest lifetimes on the equal-frequency contour shown in Fig.~\ref{Fig1RC3}(e); 
Second, the scattering channels from $\bm{k}$ to $\bm{k} e^{i \theta}$ ($\theta \neq 0, \pi$) in most directions $\theta$ are suppressed by dynamical degeneracy splitting while in several directions are reserved as we discussed in the anomalous scattering in the main text;  correspondingly, the petal shape inside the black dashed circle in Fig.~\ref{Fig1RC3}(c) disappears in Fig.~\ref{Fig1RC3}(f). 
Therefore, quasiparticle interference pattern reflects the geometric shape and the dynamical degeneracy splitting on the equal-frequency contour. 

\section{Scattering theory in non-Hermitian tight-binding model}

In this section, we first give the general form of the scattering equation. 
Then, we derive Eq.(4) of the main text and extend it into multi-band and multiple-pole cases. 
After this, we demonstrate that there are two cases for the scattered waves. 
Finally, we discuss the relation between dynamical degeneracy splitting and the pole's behavior shown in Fig.~2 of the main text. 

\subsection{The general scattering theory}

We begin with a general non-Hermitian Hamiltonian in $d$ dimensions, 
\begin{equation}\label{SM2A_GeneralHam}
	\begin{split}
		H=H_0+V  = \sum_{\bm{r}}\sum_{\bm{l}} \sum_{\alpha,\beta = 1}^q t_{\bm{l},\alpha\beta} | \bm{r}, \alpha\rangle \langle \bm{r}+\bm{l}, \beta | + \sum_{\bm{r}} \sum_{\alpha,\beta = 1}^q \mathcal{V}_{\alpha\beta}(\bm{r})| \bm{r}, \alpha\rangle \langle \bm{r}, \beta |,
	\end{split}
\end{equation}
where $H_0$ is the free Hamiltonian and $V$ represents the scattering potential. 
Here $\bm{r} = (r_1, r_2, \dots, r_d)$ represents the lattice coordinate in $d$ dimensions, and $\bm{l}=(l_1, l_2, \dots, l_d)$ indicates the hopping displacement with the largest distance $|\bm{l}_{\mathrm{max}}|$. 
$\alpha$ and $\beta$ label different degrees of freedom per unit. 
Note that the free Hamiltonian has translation symmetry, which can be written in momentum space as, 
\begin{equation}\label{SM2A_BlochHam}
	H_0 = \sum_{\bm{k}\in \mathrm{BZ}} \sum_{\bm{l}} \sum_{\alpha,\beta = 1}^q t_{\bm{l},\alpha\beta} e^{i \bm{k} \bm{l}} |\bm{k}, \alpha\rangle \langle \bm{k}, \beta| = \sum_{\bm{k}\in \mathrm{BZ}} \mathcal{H}_0(\bm{k})|\bm{k}\rangle \langle \bm{k}|,
\end{equation}
where $\mathcal{H}_0(\bm{k})$ is a $q \times q$ non-Hermitian Bloch Hamiltonian. 
Then, we define the right and left eigenstates~\cite{SM_Brody2014} as
\begin{equation}\label{SM2A_BiBlochWave}
	\begin{split}
		&\mathcal{H}_0(\bm{k})|u_{\mu}^R(\bm{k})\rangle = E_{\mu}(\bm{k})|u_{\mu}^R(\bm{k})\rangle; \\
		&\langle u_{\mu}^L(\bm{k})|\mathcal{H}_0(\bm{k}) = E_{\mu}(\bm{k})\langle u_{\mu}^L(\bm{k})|. 
	\end{split}
\end{equation}
Here we assume the Bloch Hamiltonian has no spinful reciprocity~\cite{SM_Kawabata2020} to enforce the band degeneracy, namely $E_{\nu}(\bm{k})=E_{\tilde{\nu}}(-\bm{k})$. 

Next, we establish the scattering theory in the $d$-dimensional non-Hermitian lattice. 
Once the scattering process occurs far away from the boundary, the Bloch Hamiltonian is a good starting point. 
Consider an incident wave $|\phi_i\rangle$ as an eigenstate with energy $E_0$ of the free Hamiltonian $H_0$. 
The local scattering potential $V$ is a perturbation to the continuum spectra of scattering states in the thermodynamic limit. 
Therefore, we have two eigenequations, 
\begin{align}
	&H_0|\phi_i\rangle = E_0 |\phi_i\rangle; \label{SM2A_EigenEqusA}\\ 
	(H_0+V)|\psi\rangle &= E_0 |\psi\rangle, \,\,\,\, |\psi\rangle = |\phi_i\rangle + |\phi_s\rangle. \label{SM2A_EigenEqusB}
\end{align}
Here $|\psi\rangle$ represents the eigenstate of the full Hamiltonian $H$, which comprises the incident wave $|\phi_i\rangle$ and scattering wave $|\phi_s\rangle$. 
Subtracting Eq.(\ref{SM2A_EigenEqusA}) from Eq.(\ref{SM2A_EigenEqusB}), we can get $(E_0-H_0)(|\psi\rangle - |\phi_i\rangle) = V|\psi\rangle$. 
Further, we can obtain the scattering equation, 
\begin{equation}\label{SM2A_ScatEqua}
	|\psi\rangle = |\phi_i\rangle + G_0^{+}(E_0) V |\psi\rangle,
\end{equation}
where $G_0^{+}(E_0) = [E_0 - H_0 + i \eta]^{-1}$ with $\eta \rightarrow 0^+$. 
Finally, we can write the scattering equation in real space, i.e., 
\begin{equation}\label{SM2A_RealSpaceSE}
	\begin{split}
		\psi_{\alpha}(\bm{r}) &= \phi_{i,\alpha}(\bm{r}) + 	\phi_{s,\alpha}(\bm{r}) \\
		&= \phi_{i,\alpha}(\bm{r}) +  \sum_{\beta \sigma}\int d\bm{r}^{\prime} G_{0,\alpha\beta}^{+}(E_0; \bm{r}, \bm{r}^{\prime}) \mathcal{V}_{\beta\sigma}(\bm{r}^{\prime}) \psi_{\sigma}(\bm{r}^{\prime}),
	\end{split}
\end{equation}
where $\phi_{s,\alpha}(\bm{r}) = \langle \bm{r}, \alpha| \phi_s\rangle $ is the $\alpha$-th component of scattered wave at $\bm{r}$-th lattice site. 
It is clear that the characteristics of scattered wave are encoded in the retarded Green function. 
Note that $E_0$ is an eigenvalue of Hamiltonian $H_0$, hence $E_0\in \{E | E=E_\mu(\bm{k}), \forall \mu, \bm{k}\in\mathrm{BZ}\}$. 
Without loss of generality, we assume $E_0=E_{\nu}(\bm{k}_i)$. 
The retarded Green function can be further written as  
\begin{equation}\label{SM2A_GreenFunc}
	G_{0,\alpha\beta}^+(E_{\nu}(\bm{k}_i); \bm{r},\bm{r}^{\prime}) = \int_{\mathrm{BZ}} d \bm{k}  \, e^{i \bm{k} (\bm{r}-\bm{r}^{\prime})}  \langle \alpha | [E_{\nu}(\bm{k}_i)+i\eta-\mathcal{H}(\bm{k})]^{-1}|\beta\rangle = \sum_{\mu=1}^q \int_{\mathrm{BZ}}d \bm{k}  \, e^{i \bm{k} (\bm{r}-\bm{r}^{\prime})}
	\frac{\mathcal{P}_{\mu,\alpha\beta}(\bm{k})}{E_{\nu}(\bm{k}_i)+i\eta-E_\mu(\bm{k})},
\end{equation}
where $\mathcal{P}_\mu(\bm{k}) = | u^{R}_{\mu}(\bm{k})\rangle\langle u^L_\mu(\bm{k})|$ is the projection operator on $\mu$-th non-Hermitian Bloch band, and $|u_{\mu}^{R/L}(\bm{k})\rangle$ is the right/left eigenstate of Bloch Hamiltonian $\mathcal{H}_0(\bm{k})$ defined in Eq.(\ref{SM2A_BiBlochWave}). 

\subsection{The derivation of Eq.(4) in the main text}
Here, we consider the scattering potential in $d$-dimensional non-Hermitian system with impurities of codimension-1. 
It can be expressed as 
\begin{equation}\label{SM2A_CoOneImpScat}
	\mathcal{V}_{\beta\sigma} (\bm{r}) = \lambda \, \delta_{\beta\sigma} \, \delta(r_{\bot}=0),
\end{equation}
where $\lambda$ is the strength of impurities. 
We specify these codimension-1 impurities lying on the $r_{\bot}=0$, that is point, line or plane in $d=1,2$ or $3$ dimensions, respectively. 
Therefore, the full Hamiltonian has still the translation symmetry along remaining $d-1$ directions. 
Correspondingly, the scattering equation can be written in $(\bm{k}_{\parallel}, r_{\bot})$ representation with $\bm{k}_{\parallel} = (k_1, k_2, \dots, k_{d-1})$. 
We can always expressed the momentum of the incident wave as $\bm{k}_i = (\bm{k}_i^{\parallel},k_{\bot})$. 
Combining Eq.(\ref{SM2A_RealSpaceSE}) with Eq.(\ref{SM2A_GreenFunc}), one can finally obtain the scattered wave with codimension-1 impurities as
\begin{equation}\label{SM2A_ScatWave}
	\phi_{s,\alpha}(\bm{k}_{i}^{\parallel}, r_{\bot}) = \lambda  \sum_{\mu, \sigma=1}^{q} \int_{-\pi}^{\pi} d k_{\bot} e^{i k_{\bot}r_{\bot}} \frac{\mathcal{P}_{\mu,\alpha\sigma}(\bm{k}_{i}^{\parallel},e^{i k_{\bot}})}{E_{\nu}(\bm{k}_i)+i \eta - E_{\mu}(\bm{k}_i^{\parallel},e^{i k_{\bot}})} \psi_{\sigma}(\bm{k}_{i}^{\parallel},0).
\end{equation}

Now we consider the $q=1$ case in $d=2$ dimensions, which gives the same properties of scattered wave as multi-band ($q>1$) cases, but in a more compact form. 
In this case, the codimension-1 scattering potential in Eq.(\ref{SM2A_CoOneImpScat}) becomes an impurity line lying on the $r_{\bot}=0$ and along the $\theta$-direction, which is denoted as $\mathcal{L}_{\theta}$. 
For the left (right) side of the impurity line $\mathcal{L}_\theta$, the corresponding region is denoted by $r_{\bot}<0$ ($r_{\bot}>0$), as shown in Fig.~\ref{SM_FigS2}(a). 
We define $z:=e^{i k_{\bot}}$, then the scattered wave becomes
\begin{equation}\label{SM2A_ScatWaveA}
	\phi_{s}({k}_{i}^{\theta}, r_{\bot}) =\lambda \, \psi({k}_{i}^{\theta},0) \oint_{|z|=1} \frac{d z}{z}  \frac{ z^{r_{\bot}} }{E(\bm{k}_i)+i \eta - \mathcal{H}_0({k}_{i}^{\theta},z)},
\end{equation}
where $k_{i}^{\theta}=\bm{k}_i\cdot \bm{e}_\theta$ is the $\theta$-component of the incident momentum and  $\bm{e}_\theta$ is the unit vector along $\theta$ direction in $\bm{k}$ space.  
The coefficient  $\psi(k_{i}^{\theta},0)$ is a constant for a given $k_{i}^{\theta}$, which can be solved by Eq.(\ref{SM2A_ScatWaveA}). 
Note that for given incident wave $\bm{k}_i$ and impurity line $\mathcal{L}_{\theta}$, $E(\bm{k}_i)+i \eta - \mathcal{H}_0({k}_{i}^{\theta},z) $ in Eq.(\ref{SM2A_ScatWaveA}) is a complex function of variable $z$. 
Therefore, we can define
\begin{equation}\label{SM2A_ScatWaveB}
	g(E(\bm{k}_i)+i \eta, k_i^{\theta}, z) = E(\bm{k}_i)+i \eta - \mathcal{H}_0({k}_{i}^{\theta},z) = \frac{(E(\bm{k}_i)+i \eta)z^{m} - P_{m+n}(k_i^{\theta}, z)}{z^{m}}
\end{equation}
with an $m$-order pole at the origin and $m+n$ simple zeros located in the complex $z$ plane. 
We mark those zeros inside $|z|=1$ curve as $\{z_{\mathrm{in}}\}$ and those outside $|z|=1$ as $\{z_{\mathrm{out}}\}$. 
Physically, the scattered wave is required to converge at infinity ($|r_{\bot}|\rightarrow \infty$) and be continuous at the impurity line ($r_{\bot}=0$). 
Taking these into account, the scattered wave in Eq.(\ref{SM2A_ScatWaveA}) can be finally obtained as 
\begin{equation}\label{SM2A_ScatWaveC}
	\phi_s({k_{i}^{\theta}},r_{\bot}) = \lambda \psi({k_{i}^{\theta}},0) 
	\left\{  
	\begin{split}
		&\sum_{|z_{\mathrm{in}}|<1} {\mathrm{C}}(z_{\mathrm{in}}) z_{\mathrm{in}}^{r_{\bot}}, \,\,\,\,\,\,\,\,\, r_{\bot} > 0; \\  
		&\sum_{|z_{\mathrm{out}}|>1} - {\mathrm{C}}(z_{\mathrm{out}}) z_{\mathrm{out}}^{r_{\bot}}, \,\,\, r_{\bot} < 0,
	\end{split}  
	\right.  
\end{equation}
where $\mathrm{C}(z_{\mathrm{in}/\mathrm{out}})$ equals $2\pi i$ times the residue of the function  $[z\, g(E(\bm{k}_i)+i \eta,k_i^{\theta}, z)]^{-1}$ at the pole $z_{\mathrm{in}/\mathrm{out}}$ inside/outside the $|z|=1$ curve. 
So far, we have the Eq.(4) in the main text. 

From Eq.(\ref{SM2A_ScatWaveC}) we can see that the behavior of scattered waves is dominated by those poles $z_{\mathrm{in/out}}$ when $|r_{\bot}|\rightarrow \infty$, which is the same as that in $q>1$ cases. 
Next, we will discuss the form of Eq.(4) in the multi-band and multiple pole cases. 

\subsection{Eq.(4) in the multi-band and multiple pole cases}

We begin with the Eq.(\ref{SM2A_RealSpaceSE}). 
The $\alpha$-component of the scattered wave can be expressed as $\phi_{s,\alpha}(\bm{r})=\langle \bm{r}, \alpha| \phi_s\rangle$, where $\alpha$ represents the internal degree of freedom of the unit cell. 
For simplicity, we can take the basis transformation from this natural basis to Bloch band basis, that is, 
\begin{equation}
	\phi_{s,\alpha}(\bm{r}) = \langle \bm{r}, \alpha| \phi_s\rangle = \langle \bm{r}| \otimes \sum_{\mu=1}^q \langle \alpha|u_{\mu}^R\rangle \langle u_{\mu}^L|\phi_s\rangle = \sum_{\mu=1}^q S_{\alpha \mu} \phi_{s,\mu}(\bm{r}),
\end{equation}
where $\phi_{s,\mu}(\bm{r}) = \langle \bm{r} |\otimes \langle u_{\mu}^L| \phi_s\rangle$, and $|u_{\mu}^R\rangle$ is the right wave vector on the $\mu$-th band and satisfies the bi-orthogonality and normalization. 
Once the $\mu$-th band component of the scattered wave is obtained, the scattered wave in natural basis can be obtained by the above basis transformation. 

The $\phi_{s,\mu}(\bm{r})$ can be expressed as 
\begin{equation}\label{R2_MBScatWave}
	\begin{split}
		\phi_{s,\mu}(\bm{r}) = \sum_{\nu,\xi=1}^q\int d\bm{r}^{\prime} G_{0,\mu\nu}^{+}(E_0; \bm{r}, \bm{r}^{\prime}) \mathcal{V}_{\nu\xi}(\bm{r}^{\prime}) \psi_{\xi}(\bm{r}^{\prime}),
	\end{split}
\end{equation}
with $E_0$ the energy of the incident wave. 
Correspondingly, the matrix element of the Green's operator becomes 
\begin{equation}\label{R2_MBScatWavev1}
	\begin{split}
		G_{0,\mu\nu}^{+}(E_0; \bm{r}, \bm{r}^{\prime}) 
		&= \int_{\mathrm{BZ}} d \bm{k}  \, e^{i \bm{k} (\bm{r}-\bm{r}^{\prime})}  \langle u_{\mu}^L | [E_{0}+i\eta-\mathcal{H}(\bm{k})]^{-1}|u_{\nu}^R\rangle \\
		&= \int_{\mathrm{BZ}} d \bm{k}  \, e^{i \bm{k} (\bm{r}-\bm{r}^{\prime})} \frac{\delta_{\mu\nu}}{E_0+i \eta - E_{\mu}(\bm{k})} .
	\end{split}
\end{equation}
In general, the codimension-1 scattering potential can be written as 
\begin{equation}\label{R2_ScatterPotential}
	V(\bm{r}) = \sum_{\nu,\xi=1}^{q} \lambda \, \mathcal{V}_{\nu\xi} \, \delta(r_{\bot}=0) |u_{\nu}^{R}\rangle\langle u_{\xi}^L|, 
\end{equation}
Combining Eq.(\ref{R2_MBScatWavev1}) and Eq.(\ref{R2_ScatterPotential}), we can obtain the the $\mu$-th component scattered wave as
\begin{equation}\label{SM_MBScatWavev2}
	\phi_{s,\mu}(k_i^{\parallel},r_{\bot}) = \lambda \, \sum_{\xi=1}^{q} \int \frac{dz}{z} \, z^{r_{\bot}} \, \frac{\mathcal{V}_{\mu\xi}}{E_0+i\eta-E_{\mu}(k_i^{\parallel},z)} \psi_{\xi}(k_i^{\parallel},0). 
\end{equation}
This equation shows that the general scattered potential couples the different band components to form the scattered wave. 
Note that $\mathcal{V}_{\mu\xi}$ here is the matrix representation under the Bloch basis. 
For simplicity, the scattering potential is also chosen to be 
\begin{equation}\label{R2_ScatterPotentialv1}
	V(\bm{r}) = \sum_{\nu,\xi=1}^{q} \lambda \, \delta_{\nu\xi} \, \delta(r_{\bot}=0) |u_{\nu}^{R}\rangle\langle u_{\xi}^L|, 
\end{equation}
which always has the diagonal matrix form under any basis transformation. 
Then, Eq.(\ref{SM_MBScatWavev2}) becomes 
\begin{equation}\label{R2_MBScatWavev3}
	\phi_{s,\mu}(k_i^{\parallel},r_{\bot}) = \lambda \, \psi_{\mu}(k_i^{\parallel},0) \int \frac{dz}{z} \, \frac{z^{r_{\bot}}}{E_0+i\eta-E_{\mu}(k_i^{\parallel},z)},
\end{equation}
which reduces to the one-band form as shown in Eq.(\ref{SM2A_ScatWaveA})  of the supplementary materials, and the similar form of Eq.(4) in the main text can be obtained with the band index
\begin{equation}\label{R2_MBEq4}
	\phi_{s,\mu}({k_{i}^{\theta}},r_{\bot}) = \lambda \psi_{\mu}({k_{i}^{\theta}},0) 
	\left\{  
	\begin{split}
		&\sum_{|z_{\mathrm{in}}|<1} {\mathrm{C}_{\mu}}(z_{\mathrm{in}}) z_{\mathrm{in}}^{r_{\bot}}, \,\,\,\,\,\,\,\,\, r_{\bot} > 0; \\  
		&\sum_{|z_{\mathrm{out}}|>1} - {\mathrm{C}_{\mu}}(z_{\mathrm{out}}) z_{\mathrm{out}}^{r_{\bot}}, \,\,\, r_{\bot} < 0,
	\end{split}  
	\right.  
\end{equation}
where $\mathrm{C}_{\mu}(z)$ corresponds to the residue for the $\mu$-th band. 

Next, we discuss the form of Eq.(4) in the cases having multiple pole. 
In this case, the only different thing is the calculation of the coefficient $C(z_i)$. 
We need to take the multiplicity of each pole $z_i$ under consideration. 
Consider the one-band system with the $n$-order pole $z_i$, then the coefficient $C(z_i)$ in Eq.(4) of the main text can be calculated as 
\begin{equation}\label{R2_MPCase}
	C(z_i) = 2\pi i \, \mathrm{Res} (F(z),z_i) =  \frac{2\pi i}{n!} \lim_{z\rightarrow z_i} \frac{d^{n-1}}{dz^{n-1}} [(z-z_i)^n F(z)]
\end{equation}
with $F(z) = [z g(E_0+i\eta, k_i^{\parallel}, z)]^{-1}$, and $g(E_0+i\eta, k_i^{\parallel}, z )$ is given in Eq.(\ref{SM2A_ScatWaveB}) of the supplementary materials. 

\subsection{Two cases of scattering waves}

\begin{figure}[t]
	\begin{center}
		\includegraphics[width=1\linewidth]{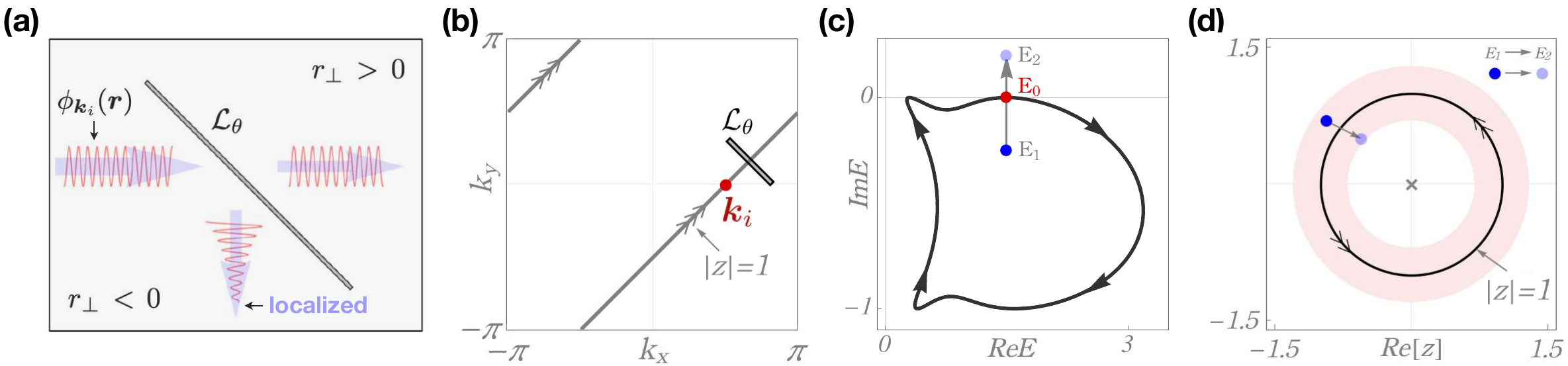}
		\par\end{center}
	\protect\caption{\label{SM_FigS2} (a) The schematic of the anomalous scattering process. (b) The path in momentum space preserves the component along $\theta$ direction and transverses the incident wave $\bm{k}_i$, which is denoted by $|z|=1$. (c) The Bloch spectrum of $\mathcal{H}(\bm{k})$ in Eq.(\ref{SM_TBModel}) as $\bm{k}$ transverses the path in (b). (d) the flow of zeros $z$ in Eq.(\ref{SM2A_ScatWaveB}) when the reference energy changes from $E_1$ to $E_2$, here only the zeros in the red annular region are shown. } 
\end{figure}

Now we demonstrate that there are only two cases of scattered wave as discussed in the main text. 
It means that there are two types of behaviors of the poles $\{z_{\mathrm{in}/\mathrm{out}}\}$ in Eq.(\ref{SM2A_ScatWaveC}). 
In case (i), there are two poles that touch the $|z|=1$ curve simultaneously from the inner and outer sides respectively when $\eta\rightarrow 0^+$; and in case (ii), there is only one pole touches the $|z|=1$ curve from the {\em inside} as $\eta\rightarrow 0^+$. 

Without loss of generality, we assume the incident plane wave $\bm{k}_i=(k_i^{\theta},k_i^{\bot})$ with energy $E_{\nu}(\bm{k}_i)$ comes from $r_{\bot}<0$ region as show in Fig.~\ref{SM_FigS2}(a). 
Correspondingly, the transmitted wave is in the $r_{\bot}>0$ region and the reflected wave is in the $r_{\bot}<0$ region. 
From Eq.(\ref{SM2A_ScatWaveC}) one can see that when $|r_{\bot}|\rightarrow \infty$, the transmitted wave is dominated by the maximum $z^{\mathrm{max}}_{\mathrm{in}} \in \{ z_{\mathrm{in}} \}$ (having the maximal amplitude $|z^{\mathrm{max}}_{\mathrm{in}} |$), and the reflected wave is mainly controlled by the minimum $z^{\mathrm{min}}_{\mathrm{out}} \in \{ z_{\mathrm{out}} \}$ (having the minimal amplitude $|z^{\mathrm{min}}_{\mathrm{out}} |$). 
Also, as can be seen from Eq.(\ref{SM2A_ScatWaveA}), when $\eta \rightarrow 0^+$ there is always a pole $z=e^{i k_i^{\bot}}$ touching $|z|=1$ curve.
Next, we show that this pole always approaches $|z|=1$ curve from the inside, not from the outside. 

Generally, for the scattering process shown in Fig.~(\ref{SM_FigS2})(a), the momentum $k_{\theta}$ along the impurity line $\mathcal{L}_{\theta}$ is preserved. 
When we give the momentum $\bm{k}_i$ of the incident wave, the path in $\bm{k}$ space that is perpendicular to the $\theta$ direction is determined. 
This path traverses the momentum $\bm{k}_i$, as shown in Fig.~(\ref{SM_FigS2})(b). 
Here we use the Hamiltonian in Eq.(\ref{SM_TBModel}) as an example, and choose the incident wave to be $\bm{k}_{i} = (k_i^x,k_i^y) = (\pi/2,0)$ with the energy $E_0=3/2$. 
As $\bm{k}$ transverses the path in Fig.~(\ref{SM_FigS2})(b), a loop-shape spectrum on the complex energy plane can be obtained as plotted in Fig.~(\ref{SM_FigS2})(c). 
Note that the incident wave has the rightward velocity, which requires $v_x(\bm{k_i})>0$ defined in Eq.(\ref{SM_GroupVelocity}). 

Now we can define the Bloch spectral winding number regarding the reference energy $E_0+i \eta$~\cite{SM_Kai2020,SM_Okuma2020} as 
\begin{equation}\label{SM2A_SpecWinding}
	w(E_0+i \eta) =\frac{1}{2\pi} \oint_{|z|=1} \frac{d}{dz} \mathrm{arg} \, [g(E_0+i \eta, k_i^{\theta}, z)] dz = N_{\mathrm{zeros}} -  N_{\mathrm{poles}}, 
\end{equation}
where $N_{\mathrm{zeros}}$ refers to the number of zeros of $g(E_0+i \eta, k_i^{\theta}, z)$ inside $|z|=1$ and $N_{\mathrm{poles}}$ is the order of the pole at the origin. 
The spectral winding number is ill-defined when $\eta=0$. 
As shown in Fig.~\ref{SM_FigS2}(c), the $\eta$ is tuned from $\eta<0$ to $\eta>0$ (correspondingly, the reference energy goes from $E_1\rightarrow E_0\rightarrow E_2$). 

Note that we always choose the rightward incident wave as shown in Fig.~\ref{SM_FigS2}(a).
Correspondingly, as $k$ increases, the Bloch spectrum always passes through $E_0$ from right to left in the complex energy plane, as shown in Fig.~\ref{SM_FigS2}(c). 
Therefore, the spectral winding number always increases by $1$ when the reference energy $E_0+i \eta$ runs from the region below $E_0$ to the region above $E_0$, namely $w{(E_0+0^+)}-w{(E_0+0^-)}=1$. 
For example, in Fig.~\ref{SM_FigS2}(c), $w{(E_1)}=-1$ and $w{(E_2)}=0$. 
Meanwhile, in this process $N_{\mathrm{poles}}$ is always invariant, thus the number of zeros inside $|z|=1$ curve $N_{\mathrm{zeros}}$ increases $+1$. 
Equivalently, there is one pole in Eq.(\ref{SM2A_ScatWaveC}) that moves into $|z|=1$ curve as the reference energy changes from $E_1$ to $E_2$, as shown in Fig.~\ref{SM_FigS2}(d). 
Therefore, for the rightward incident wave, there is always one pole approaches to $|z|=1$ curve from inside when $\eta\rightarrow 0^+$. 
Finally, we can obtain only the two cases of scattered wave as discussed in the main text. 

\subsection{The relation between dynamical degeneracy splitting and pole's behavior}

Here, we demonstrate that when dynamical degeneracy splitting occurs, there are at least two poles in Eq.(\ref{SM2A_ScatWaveB}) reach $|z|=1$; when dynamical degeneracy splitting does not occur, only one pole approaches $|z|=1$, as illustrated in Fig.~2 of the main text. 

For simplicity, one-band case is considered below. 
Assuming that the impurity line is along the $\theta$ direction (labeled as $\mathcal{L}_{\theta}$) and the incident wave has (complex) energy $E_0=\omega_0+i\Gamma_0$. 
Here, we clarify that when dynamical degeneracy splitting does not occur at $\omega_0$, for each $\mathcal{L}_{\theta}$, there are at least two poles going to $|z|=1$ as $\eta\rightarrow 0^+$; when dynamical degeneracy splitting occurs at $\omega_0$, there exist $\mathcal{L}_{\theta}$ such that only one pole approaches $|z|=1$ when $\eta\rightarrow 0^+$. 

When dynamical degeneracy splitting does not occur at $\omega_0$, as shown in Fig.~1(a)(b) of the main text, all momenta on the equal-frequency contour at $\omega_0$ have the same real part $\omega_0$ (by the definition of equal-frequency contour in Eq.(1)) and the same imaginary part $\Gamma_0$. 
Therefore, it means in this case that the solution of $E_0-H(\bm{k})=0$ is exactly the equal-frequency contour at $\omega_0$. 
Note that all momenta on the equal-frequency contour are real number. 

Once incident wave and scattering impurity line $\mathcal{L}_{\theta}$  are given, the $k_i^{\theta}$ and $E_0$ are determined, where $k_i^{\theta}$ represents the $\theta$-component of the incident wave momentum $\bm{k}_i$. 
When dynamical degeneracy splitting does not occur at $\omega_0$, for each $k_i^{\theta}$, one can always find at least two real momenta, $\bm{k}_i=(k_i^{\theta},k_{i}^{\bot})$ and $\bm{k}_s=(k_i^{\theta},k_{s}^{\bot})$ that satisfy $E_0-H(\bm{k}_i)=0$ and $E_0-H(\bm{k}_s)=0$. 
Note that $\bm{k}_i$ must belong to the solutions of $E_0-H(\bm{k})=0$ because $E_0$ is the energy of incident wave $\bm{k}_i$. 

Here, we look at the poles of 
\begin{equation}\label{R2_PolesEqua}
	\lim_{\eta\rightarrow 0^+}z [E_0+i \eta - H(k_i^{\theta},z)]^{-1} = \frac{z^{m-1}}{E_0 z^{m}-P_{m+n}(k_i^{\theta},z)},
\end{equation}
where $P_{m+n}(k_i^{\theta},z)$ is a polynomial of $z$ for given $k_i^{\theta}$. 
The poles are exactly the solutions of $E_0-H(k_i^{\theta},z)=0$ for given $k_i^{\theta}$ and $E_0$. 
Therefore, when dynamical degeneracy splitting does not occur at $\omega_0$, for each $k_i^\theta$, we have at least two solutions, $z_i=e^{i k_i^{\bot}}$ and $z_s=e^{i k_s^{\bot}}$ that satisfy $|z_i|=|z_s|=1$ because $k_i^{\bot}$ and $k_s^{\bot}$ are real. 
In other words, when dynamical degeneracy splitting does not occur at $\omega_0$, for each $k_i^\theta$, there are at least two poles approaching $|z|=1$ as $\eta\rightarrow 0^+$. 

When dynamical degeneracy splitting occurs at $\omega_0$, different momenta on the equal-frequency contour have the same real part $\omega_0$ but different  imaginary parts. 
Therefore, there exist some $k_i^{\theta}$ such that the solutions of $E_0-H(\bm{k})=0$ have only one real momentum $\bm{k}_i$ and other complex momenta, for example, $\bm{k}_s = (k_i^{\theta},k_{s}^{\bot})$ with $k_i^{\theta}\in \mathbb{R}$ but $k_s^{\bot}\in \mathbb{C}$. 
Correspondingly, the poles of Eq.(\ref{R2_PolesEqua}) include one $z_i=e^{i k_i^{\bot}}$ on the $|z|=1$ unit circle and other $z_s=e^{i k_s^{\bot}}$ with $|z_s|\neq 1$. 
Therefore, when dynamical degeneracy splitting occurs at $\omega_0$, there exist $k_i^{\theta}$ such that only one pole approaches $|z|=1$ when $\eta\rightarrow 0^+$. 

\section{Scattered waves in time evolution}

In this section, we will study the time evolution of incident waves and discuss under what conditions the stationary scattering equation (Eq.(2) in the main text) in non-Hermitian systems can describe the scattered waves well. 

The total Hamiltonian comprises the unperturbed Hamiltonian $H_0$ and scattering potential $\lambda V$ with the strength $\lambda$, that is $H=H_0+\lambda V$. 
We define their eigenequations as follows:
\begin{equation}\label{R2_EigenEquas}
	\begin{split}
		&H|\psi^R_n\rangle = E_n |\psi^R_n\rangle, \,\,\,\, 
		H^{\dagger} | \psi^L_n \rangle = E^{\ast}_n | \psi^L_n \rangle;  \\ 
		&H_0|\phi^R_n\rangle = \epsilon_n |\phi^R_n\rangle, \,\,\,\, 
		H_0^{\dagger} | \phi^L_n \rangle= \epsilon^{\ast}_n | \phi^L_n \rangle,
	\end{split}
\end{equation}
where the superscript $R$ and $L$ represent the right and left eigenvectors, respectively. 
They satisfy the bi-orthogonality, namely, $\langle \psi^L_n| \psi_m^R\rangle = \langle \phi^L_n| \phi_m^R\rangle = \delta_{mn}$. 
The incident wave (initial state) is assumed to be an eigenstate $|\phi_0\rangle$ of $H_0$ with the energy $\epsilon_0$. 
After encountering the scattering potential, the time evolution of the scattered wave can be defined as: 
\begin{equation}\begin{aligned}\label{R2_TimeSactterWave}
		|\phi_s(t)\rangle = U(t)|\phi_0\rangle - U_0(t)|\phi_0\rangle, 
\end{aligned}\end{equation}
where the time evolution operators reads $U(t)=e^{-i H t}$ and $U_0(t)=e^{-i H_0 t}$. 
Under spectral representation, the above equation can be further expressed as 
\begin{equation}\begin{aligned}\label{R2_TimeSactterWave_1}
		|\phi_s(t)\rangle &= \sum_me^{-iE_mt}|\psi_m^R\rangle\langle \psi_m^L|\phi_0\rangle-e^{-iE_0t}|\phi_0\rangle\\
		&= e^{- i E_0 t} [ \langle \psi^L_0|\phi_0\rangle |\psi_0^R\rangle - |\phi_0\rangle 
		+ \sum_{m\neq 0} e^{-i (E_m-E_0) t}  \langle \psi^L_m|\phi_0\rangle |\psi_m^R\rangle ]. 
\end{aligned}\end{equation}
Here, $\langle \psi_0^L|$ is the left eigenvector of $H$ with the same complex energy of the initial state $|\phi_0\rangle$, that is $E_0=\epsilon_0$. 
From this one can see that when (i) $\langle \psi^L_0|\phi_0\rangle \approx 1$ and (ii) $\sum_{m\neq 0} e^{-i (E_m-E_0) t}  \langle \psi^L_m|\phi_0\rangle |\psi_m^R\rangle\simeq 0$, the scattered wave in Eq.(\ref{R2_TimeSactterWave_1}) reduces to 
\begin{equation}\label{R2_TimeSactterWave_2}
	|\phi_s(t)\rangle \approx e^{- i E_0 t} (|\psi_0^R\rangle - |\phi^R_0\rangle),
\end{equation}
which is exactly the time evolution form given by the Lippmann-Schwinger equation. 
Therefore, if the approximations (i) and (ii) hold, the Lippmann-Schwinger equation in Eq.(2) of the main text gives the scattered waves for the non-Hermitian Hamiltonian. 

Next, we discuss the approximation conditions under which Eq.(\ref{R2_TimeSactterWave_2}) holds. 
We divide the discussion by the following two cases: 

\begin{itemize}
	\item The incident wave has the largest imaginary energy (the lowest decay rate).
	
	In this case, $\Im{\epsilon_m}<\Im{\epsilon_0}$ for all $\epsilon_m\neq \epsilon_0$, where $\epsilon_0$ is the energy of the incident wave. 
	Note that in the scattering process of a the impurity line, the momentum $k_0^{\parallel}$ of the incident wave, which is parallel to the impurity line, is conserved. 
	Therefore, the largest imaginary energy here refers to the spectrum of the reduced 1D Hamiltonian $H(k_0^{\parallel},k^{\perp})$, instead of the whole spectrum of $H(\bm{k})$. 
	
	In what follows, we will use the perturbation theory to calculate $E_n$, $\langle \psi_0^L|$,  and $\langle \psi_m^L|$. 
	
	{\em Eigenvalue $E_n$}: With the first-order correction, the $n$-th eigenvalue of $H$ can be approximated to
	\begin{equation}\label{R2_SpectrumAppro}
		E_n \approx \epsilon_n + \langle \phi_n^L| \lambda V |\phi_n^R\rangle
		= \epsilon_n + O\left(\frac{\lambda V_0}{N}\right),
	\end{equation}
	where $|\phi_n^R\rangle$ and $\langle\phi_n^L|$ are the extended waves (plane waves) and satisfy normalization condition $\langle \phi_n^L|\phi_n^R\rangle=1$, thus each element $\langle r|\phi_n^R\rangle$ has the order of $1/\sqrt{N}$ with $N$ representing the system size. 
	Given that $\lambda V$ is a local operator in the real space, therefore, we have the conclusion: $\langle \phi_n^L| \lambda V |\phi_n^R\rangle$ has the order of $\lambda V_0/N$, where $V_0$ represents the characteristic energy of the scattering potential.  
	It tells us that under the thermodynamic limit, the continuum spectrum of $H$ well approximates to that of $H_0$. 
	
	{\em Eigenstates $\langle \psi_0^L|$,  and $\langle \psi_m^L|$}: Here, the left eigenvector of $H$ can be obtained with the first-order correction: 
	\begin{equation}\label{R2_WaveAppro}
		\langle \psi_n^L| \approx \langle \phi_n^L| + \sum_{m\neq n} \frac{\langle \phi_n^L| \lambda V| \phi_m^R\rangle}{\epsilon_m-\epsilon_n} \langle \phi_m^L|,
	\end{equation}
	where the minimal value of $\epsilon_m-\epsilon_n$ can be approximated by $E_{bw}/N$ with $E_{bw}$ representing the bandwidth of the system. Therefore, 
	\begin{equation}
		\langle \psi_n^L| \approx \langle \phi_n^L| + \sum_{m\neq n} O\left(\frac{\lambda V_0}{E_{bw}}\right) \langle \phi_m^L|.
	\end{equation}
	Note that the perturbation condition requires that the off-diagonal term $(\lambda V_0)/E_{bw} \ll 1$. 
	Therefore, one can obtain the following approximated bi-orthogonal normalization condition, 
	\begin{equation}\label{bi-orthogonal normalization}
		\langle \psi^L_m|\phi^R_0\rangle\approx\delta_{m0}+O\left(\frac{\lambda V_0}{E_{bw}}\right)
	\end{equation}
	Note that in the Eq.(\ref{R2_TimeSactterWave_1}), $e^{-i (E_m-E_0) t} $ will decay with time due to face that $\Im{\epsilon_m}<\Im{\epsilon_0}$. Therefore, putting Eq.(\ref{bi-orthogonal normalization}) into Eq.(\ref{R2_TimeSactterWave_1}), one can obtain 
	\begin{equation}\label{R2_ScatteredWaveAppro}
		|\phi_s(t)\rangle \approx e^{- i E_0 t} ( |\psi_0^R\rangle - |\phi^R_0\rangle ),
	\end{equation}
	which shows that the Lippmann-Schwinger equation in Eq.(2) can capture the scattered wave in the non-Hermitian system. 
	
	Actually, we can show that the adiabatic theorem still holds for the eigenstates with largest imaginary energy, which is another way to justify that Eq.(2) indeed gives the scattered states for non-Hermitian systems. 
	We will discuss the adiabatic theorem for the eigenstates with largest imaginary energy later. 
	
	\item The incident wave has no the largest imaginary energy.
	
	In this case,  $e^{-i (E_m-E_0) t} $ will become larger and larger than $1$ if  $\Im{\epsilon_m}>\Im{\epsilon_0}$. Therefore the approximation in Eq.(\ref{R2_TimeSactterWave_2}) holds only when the evolution time $t \ll t_c$, where $t_c$ represents the characteristic time. 
	It means that for this case the Eq.(2) can capture the scattered waves with short-time evolution. 
	This makes sense because after a long time, the eigenstates with the largest imaginary energy will dominate the dynamics of the system. 
	
	Now we examine the characteristic time $t_c$. From $e^{-i (E_m-E_0) t}  \langle \psi^L_m|\phi_0\rangle |\psi_m^R\rangle$, one can find that if $e^{-i (E_m-E_0) t}  \langle \psi^L_m|\phi_0\rangle\ll 1,$ 
	the Eq.(\ref{R2_TimeSactterWave_2}) still holds. Therefore, the characteristic time can be given by
	\begin{equation}
		e^{-i (E_m-E_0) t_c}  \langle \psi^L_m|\phi_0\rangle=e^{-i (E_m-E_0) t_c}  \frac{\langle \phi_m^L| \lambda V| \phi_0^R\rangle}{\epsilon_0-\epsilon_m}= 1,
	\end{equation}
	
	Based on the above perturbation analysis in Eq.(\ref{R2_SpectrumAppro}) and Eq.(\ref{R2_WaveAppro}), we know that
	\begin{equation}
		c_{n0}:=\langle \psi^L_{n\neq 0}|\phi^R_0\rangle\approx \frac{\langle \phi_n^L| \lambda V| \phi_0^R\rangle}{\epsilon_n-\epsilon_0}; \,\,\,\, c_{00}:=\langle \psi^L_0|\phi^R_0\rangle \approx 1. 
	\end{equation}
	Therefore, the scattered wave in Eq.(\ref{R2_TimeSactterWave_1}) can be approximated into
	\begin{equation}
		|\phi_s(t)\rangle \approx e^{- i E_0 t} \{ |\psi_0^R\rangle - |\phi^R_0\rangle 
		+ \sum_{n\neq 0} e^{-i (\epsilon_n-\epsilon_0) t}  c_{n0} |\psi_n^R\rangle \}. 
	\end{equation}
	The Eq.(\ref{R2_TimeSactterWave_2}) holds when $|e^{-i (\epsilon_n-\epsilon_0) t}  c_{n0}| \ll 1$. 
	Therefore, the characteristic time can be estimated by 
	\begin{equation}
		e^{\Delta_{n0} t_c} \frac{\lambda}{N \Delta_{n0}} \sim 1; \,\,\,\, \Rightarrow \,\,\,\,
		t_c \sim \frac{1}{\Delta_{n0}} \ln{\frac{N \Delta_{n0}}{\lambda}}, 
	\end{equation}
	where $\Delta_{n0}:=|\epsilon_n-\epsilon_0|$ represents the spectral distance between $\epsilon_n$ and $\epsilon_0$ in the complex energy plane. 
	There are two different cases: 
	
	(i) when $\Delta_{m0}$ has the order of $O(1)$ (the energies away from $\epsilon_0$), then Eq.(\ref{R2_TimeSactterWave_2}) holds when the evolution time 
	\begin{equation}\label{key}
		t \ll t_c \sim \ln{\frac{N}{\lambda}}; 
	\end{equation}
	(ii) when $\Delta_{m0}$ has the order of $1/N$, the nearest energies from the $\epsilon_0$ has the gap $\delta \epsilon \sim E_{bw}/N$, where $E_{bw}$ represents the bandwidth of the system as discussed before.
	Therefore, the Eq.(\ref{R2_TimeSactterWave_2}) holds when the evolution time 
	\begin{equation}\label{key}
		t \ll t_c \sim \frac{N}{E_{bw}} \ln{\frac{E_{bw}}{\lambda}},
	\end{equation}
	where $\lambda\ll 1$ is required such that the perturbation analysis of the wave functions can be applied. 
\end{itemize}
So far, we have clarify the applicable conditions under which the Lippmann-Schwinger equation in Eq.(2) can well describe the scattered waves.

\section{Directional invisibility and geometry-dependent skin effect}

In this section, we use a single-band example having GDSE to show the directional invisibility in terms of wave packet dynamics. 
Then, we show the transmitted and reflected intensity without time-dependent normalization factor using the two-band example (Eq.(5) in the main text). 
Finally, we prove that the normalization of spectral function still works for the dissipative non-Hermitian systems. 

\begin{figure}[t]
	\begin{center}
		\includegraphics[width=1\linewidth]{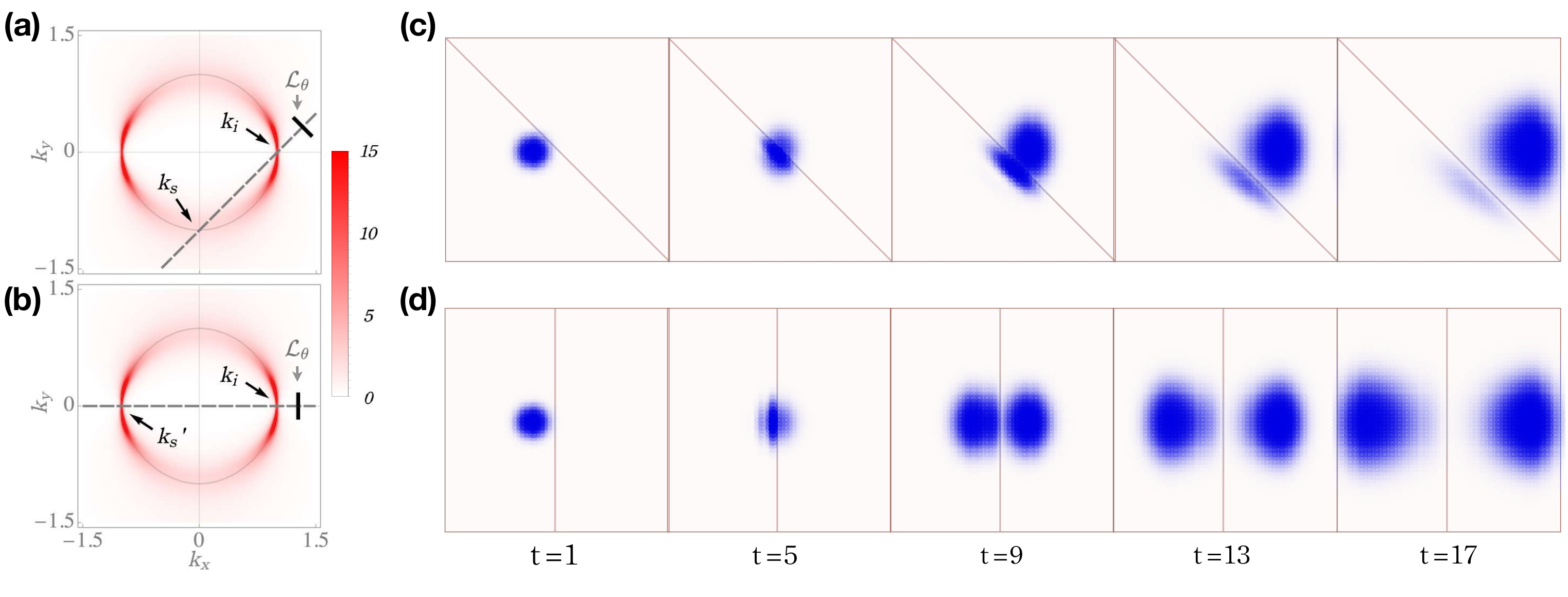}
		\par\end{center}
	\protect\caption{\label{SM_FigS3} 
		Directional invisibility in Hamiltonian Eq.(\ref{SM_SingleHam}). (a-b) show the spectral function $A(\omega,\bm{k})$ with $\omega=\mathrm{Re}\mathcal{H}(\bm{k}_i)$, where $\bm{k}_i=(k_i^x,k_i^y)=(1,0)$. 
		The intensity of spectral function corresponds to the opacity indicated in the color bar. 
		The incident wave packet with the momentum center at $\bm{k}_i$ hits the oblique impurity line in (c) and vertical impurity line in (d), where the impurity strength $\lambda=3/2$. The snapshots of wave packets at $t=1,,5,9,13,17$ are shown in the (c)(d). }
\end{figure}

\subsection{The directional invisibility in single-band model}
The Hamiltonian of the model is 
\begin{equation}\label{SM_SingleHam}
	\mathcal{H}(\bm{k}) = -2 (\cos{k_x} + \cos{k_y} )+ i g ( \cos{k_y} - 1 )
\end{equation}
with a non-Hermitian parameter $g=1$. 
Obviously, the imaginary part is $\bm{k}$-dependent, thus leading to dynamical degeneracy splitting in this system. 
We plot the spectral function $A(\omega_0,\bm{k}) = -\mathrm{Im} [G^+_0(\omega_0,\bm{k})]$ in Fig.~\ref{SM_FigS3}~(a)(b), showing the uneven broadening of equal-frequency contour (the gray circle), which is a definite signature of the occurrence of dynamical degeneracy splitting.  
According to the anomalous scattering theory, the anomalous scattering will occur. 
We assume an incident plane wave has $\bm{k}_i=(k_{x},k_{y})=(1,0)$ and hits the impurity line $\mathcal{L}_{\theta}$ with a rightward velocity in real space. 
Note that $\bm{k}_i$ lies on the equal-frequency contour $\bm{K}(\omega_0)$ with $\omega_0 = \mathrm{Re}[\mathcal{H}(\bm{k}_i)]$, as shown in Fig.~\ref{SM_FigS3}(a)(b). 
The impurity line $\mathcal{L}_{\theta}$ preserves the momentum along this direction, which means the scatterer $\mathcal{L}_{\theta}$ relates $\bm{k}_i^0$ with $\bm{k}_{s}$ and $\bm{k}_{s}^{\prime}$ in the way illustrated in Fig.~\ref{SM_FigS3}(a) and (b), respectively. 
Due to the larger broadening at $\bm{k}_{s}$, it means that the reflected wave is a non-Bloch wave, damped away from the impurity line. 
However, if we rotate the oblique impurity line into vertical as shown in Fig.~\ref{SM_FigS3}(b), the $k_y$-component will be preserved during the scattering process. 
Because of the Hamiltonian respecting $\mathcal{M}_x$ symmetry, $\bm{k}_{s}^{\prime}=\mathcal{M}^{-1}_x \bm{k}_i$ has the equal broadening with $\bm{k}_i$. 
Therefore, the vertical impurity line scatters the incident plane wave $\bm{k}_i$ to a propagating plane wave $\bm{k}_{s}^{\prime}$, then the normal scattering occurs.  
Such a phenomenon unique to GDSE is dubbed {\em directional invisibility}. 

We use the wave packet dynamics to probe the directional invisibility in this example. 
The incident wave is a Gaussian wave packet with momentum center at $\bm{k}_i=(k_i^x,k_i^y) = (1,0)$. 
The time evolution of wave packet follows $|\psi(t)\rangle = \mathcal{N}(t)e^{- i H t} |\phi_0\rangle$, where $H$ is the full Hamiltonian consisting of the free Hamiltonian in Eq.(\ref{SM_SingleHam}) and impurity-line scattering potential $\mathcal{V}(\bm{r})=\lambda \delta(r_{\bot}=0)$, and $\mathcal{N}(t)$ is the normalization factor at every time.  
The incident Gaussian wave packet has the form $\phi_0(\bm{r}) = \mathrm{exp}[-(\bm{r}-\bm{r}_0)^2/\sigma^2+i \bm{k}_i^0 \bm{r}]$. 
In Fig.~\ref{SM_FigS3}(c)(d), the parameters are set as $(x_0,y_0,\sigma)=(24,30,4)$, and the system size of the lattice is $L_x=L_y=60$. 
It can be observed that the Gaussian wave packet is almost completely transmitted through the oblique impurity line without any propagating reflected waves, as shown in Fig.~\ref{SM_FigS3}(c). 
However, parts of the wave packet are reflected by the vertical impurity line as a propagating wave, as shown in Fig.~\ref{SM_FigS3}(d). 
Therefore, the real space dynamics in Fig.~\ref{SM_FigS3}(c)(d) shows the directional invisibility. 

\subsection{The dissipative reflected and transmitted intensity}

\begin{figure}[b]
	\begin{center}
		\includegraphics[width=.8\linewidth]{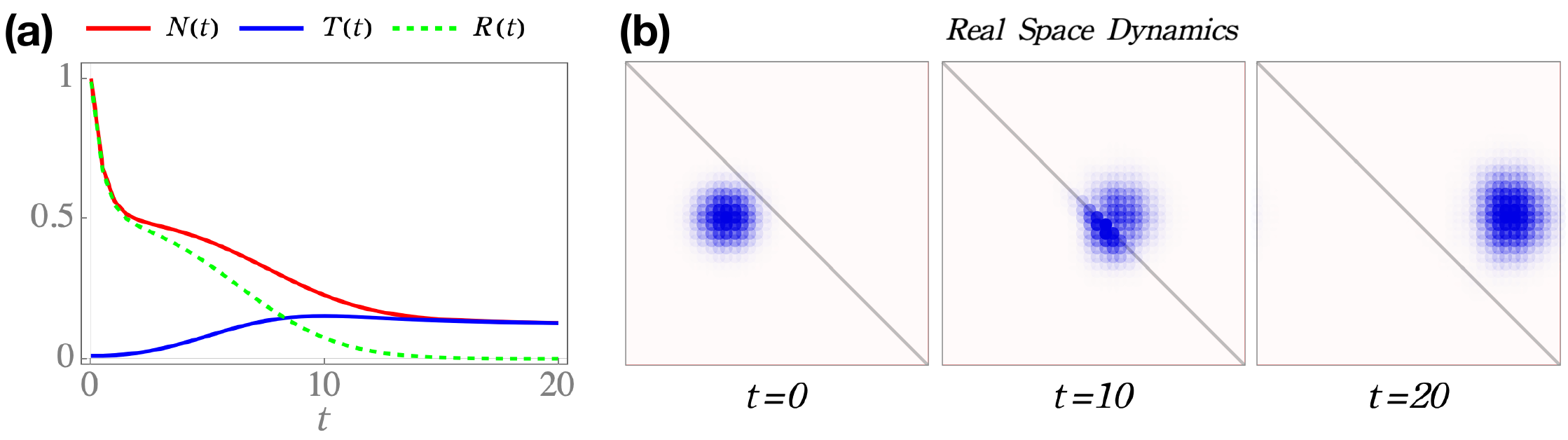}
		\par\end{center}
	\protect\caption{\label{SM_FigS4} (a) The total probability $N(t)$, transmitted wave intensity $T(t)$ and reflected wave intensity $R(t)$ change with time. Note that the wave function here is not renormalized. 
		(b) The snapshots of renormalized wave packet at  $t=0$, $t=10$, and $t=20$. }
\end{figure}

Here, we plot the intensity of transmitted and reflected components of the Gaussian wave packet in Fig.~\ref{SM_FigS4}(a). The model Hamiltonian is taken as Eq.(5) in the main text. 
The wave function has not been renormalized after every time step. 
In such a dissipative non-Hermitian Hamiltonian, the system waves always decay with time. 
Even so, it clearly shows that the Gaussian wave packet is fully transmitted. 

In our dissipative model, the Gaussian wave packet decays with time and follows the time-evolution equation: 
\begin{equation}\label{key}
	|\psi(t)\rangle = e^{-i H t}|\phi_0\rangle,
\end{equation} 
where $\phi_{0}(\bm{r})$ is the Gaussian wave packet and has the same form as that in Fig.~3 of the main text. 
The total probability of wave function at time $t$ becomes
\begin{equation}\label{key}
	N(t) = \langle \psi(t)| \psi(t)\rangle.
\end{equation}
And the transmitted and reflected intensities at time $t$ can be defined as  
\begin{equation}\label{key}
	T(t) = \sum_{\bm{r}\in S_T} \langle \psi(t)|\bm{r}\rangle \langle \bm{r} | \psi(t)\rangle; \,\,\,\, 
	R(t) = \sum_{\bm{r}\in S_R} \langle \psi(t)|\bm{r}\rangle \langle \bm{r} | \psi(t)\rangle,
\end{equation}
where $S_T$ represents the region of transmitted wave, and $S_R$ indicates the reflected region. 
In our scattering setting of the main text, $r_{\bot}>0$ is $S_T$, and $r_{\bot}<0$ is $R_T$. 
Obviously, when the scattered wave packet moves away from the scattering section (the impurity line), $T(t)+R(t)=N(t)$. 

As shown in Fig.~\ref{SM_FigS4}(a), the red solid curve represents $N(t)$ and the blue solid curve (green dotted line) corresponds to the transmitted (reflected) part. 
At the beginning, the wave packet is in the reflection region $S_R$. 
With the occurrence of scattering event, the transmitted component is greater than the reflected one, and eventually all the wave packets are in the transmitted region (correspondingly, the red and blue curve coincide after $t=15$). 
As a comparison, we select the snapshots of wave packet at $t=0$, $t=10$, and $t=20$ in Fig.~\ref{SM_FigS4}(b), where the wave packet has been renormalized as in the main text. 

\subsection{The normalization of spectral function in dissipative system}

In this part, we will prove that in the dissipative non-Hermitian systems, the normalization of the spectral function is still true for any given $\bm{k}$, that is, 
\begin{equation}
	\frac{1}{2\pi}\int d\omega A(\omega,\bm{k}) = 1.
\end{equation}
Here, the spectral function can be expressed as 
\begin{equation}
	A(\omega, \bm{k}) = -\frac{1}{N} \mathrm{Im} \mathrm{Tr} [1/(\omega + i \eta - H(\bm{k}))] = - \frac{1}{N} \mathrm{Im} \left[\sum_{\mu=1}^{N} \frac{1}{\omega+i\eta - E_{\mu}(\bm{k})} \right],
\end{equation}
where $\omega$ is real excitation frequency and $\eta=0^+$, and $\mu$ represents the band index and $N$ is the total number of energy bands. 

Note that in the dissipative systems, the imaginary part of energy satisfies $\mathrm{Im}E_{\mu}(\bm{k}) \leq 0$ for all $\mu$ and $\bm{k}$, which is considered throughout our paper. 
Now we prove the universal normalization. 
We have
\begin{equation}\label{SM_SFNormalization}
	\frac{1}{2\pi}\int_{-\infty}^{\infty} d\omega A(\omega, \bm{k}) = -\frac{1}{2\pi N} \mathrm{Im} \sum_{\mu=1}^N \int_{-\infty}^{\infty} \frac{1}{\omega+i\eta - E_{\mu}(\bm{k})} d\omega.
\end{equation}
Note that for any given band $\mu$ and momentum $\bm{k}$, the energy $E_{\mu}(\bm{k})-i \eta$ is below the real axis in the complex energy plane due to $\mathrm{Im}[E_{\mu}(\bm{k})-i \eta]<0$. 
Therefore, we use the contour integral technique, and then Eq.(\ref{SM_SFNormalization}) can be calculated as 
\begin{equation}
	\frac{1}{2\pi}\int_{-\infty}^{\infty} d\omega A(\omega, \bm{k}) = \frac{1}{2\pi N} \mathrm{Im} \sum_{\mu=1}^N \oint_{\Gamma^+} \frac{1}{\omega-[ E_{\mu}(\bm{k}) - i\eta ]} d\omega = -\frac{1}{2\pi N} \mathrm{Im}[N\times (-2\pi i )] = 1.
\end{equation}
Here, $\Gamma^+$ indicates the positively oriented integral contour that surrounds the lower half energy plane. 
Therefore, the universal normalization $\frac{1}{2\pi}\int d\omega A(\omega,\bm{k}) = 1$ is always valid for different $\bm{k}$ in the dissipative non-Hermitian systems. 

\section{The robustness of anomalous scattering}

On the theoretical perspective, we demonstrate the relation between dynamical degeneracy splitting and anomalous scattering on the general scattering potentials in the first part, which suggests the robustness of anomalous scattering (including robustness of directional invisibility in GDSE). 
Then, we numerically show that directional invisibility is robust against the changes in orbital components, and can be still observed with finite step potential and open boundary. 

\begin{figure}[b]
	\begin{center}
		\includegraphics[width=.8\linewidth]{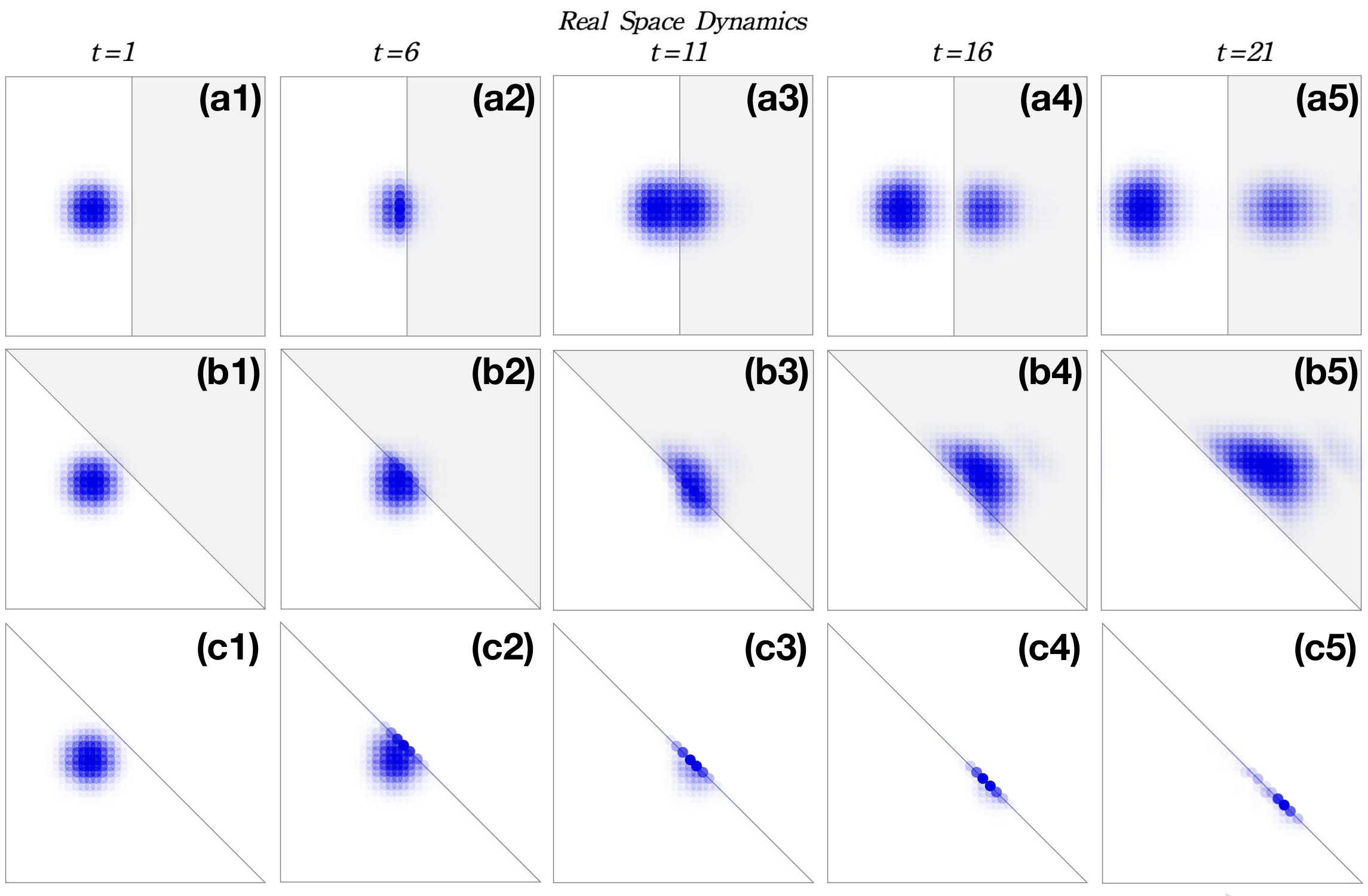}
		\par\end{center}
	\protect\caption{\label{SM_FigS5} 
		Directional invisibility in step potential(a)(b) and open boundary (c). 
		The Hamiltonian adopts Eq.(5) in the main text, and the parameters are set to be $(\mu_0,\mu_z, t_0, t, t_z, \gamma) = (1.35, -0.05, -0.4, 0.4, -0.6, 1)$.
		In (a)(b), the gray region represent the region where the step potential is nonzero ($\lambda=1$). 
		In (c), the incident wave packet hits the oblique open boundary and gets stuck, which indicates the anomalous scattering on the open boundary. }
\end{figure}

\subsection{Anomalous scattering on general scattering potential}

We begin with the general scattering equation, that is, 
\begin{equation}\label{R2_GeneralScattering}
	|\psi\rangle = |\phi_i\rangle + G_0^{+}(E_0)V|\psi\rangle 
	= |\phi_i\rangle + G_0^{+}(E_0)V|\phi_i\rangle + G_0^{+}(E_0)VG_0^{+}(E_0)V|\phi_i\rangle  + \cdots,
\end{equation}
where $|\phi_i\rangle$ represents incident wave vector with the energy $E_0$ in $H_0$. 
For analytical convenience, we here consider $V$ as a weak scattering potential and take the Born approximation, under the real space representation, then the scattering equation can be expressed as 
\begin{equation}\label{R2_WeakScatWave}
	\begin{split}
		\psi(\bm{r}) & = \phi_i(\bm{r}) + \phi_s(\bm{r})  \\ 
		& \approx \phi_i(\bm{r}) + \int d\bm{r}^{\prime} G_0^{+}(E_0; \bm{r},\bm{r}^{\prime}) \mathcal{V}(\bm{r}^{\prime})\phi_i(\bm{r}^{\prime}),
	\end{split}
\end{equation}
where $G_0^{+}(E; \bm{r},\bm{r}^{\prime})$ is the retarded free Green's function and $\mathcal{V}(\bm{r}^{\prime})$ has translation symmetry along $\theta$ direction in real space and are local function along the perpendicular direction. 
Therefore, the scattering process preserves the momentum along $\theta$ direction. 
After Fourier transformation, we have $k_{\theta}$-component scattered wave 
\begin{equation}\label{R2_thetaScatWave}
	\phi_s(k_{\theta},r_{\bot}) = \int dr^{\prime}_{\bot} G^+_0(E_0; k_{\theta}, r_{\bot}-r^{\prime}_{\bot}) \mathcal{V}(k_{\theta},r^{\prime}_{\bot})  \phi_i(k_{\theta},r^{\prime}_{\bot}),
\end{equation}
which is very similar to the results for delta potential scattering potential but scattering potential is a local function of $r_{\bot}$ in general cases. 

The scattering potential is localized along $r_{\bot}$ direction, which means
\begin{equation}\label{R2_LocalScatPotential}
	\mathcal{V}(k_{\theta},r_{\bot}) = 0 , \,\,\,\, \mathrm{if} \,\, r_{\bot} \notin S; \,\,\,\,\, S:=[r^{L}_{\bot},r^{R}_{\bot}],
\end{equation}
where $r^{L}_{\bot}$ and $r^{R}_{\bot}$ represent the left and right boundary position in $r_{\bot}$ direction. 
Therefore, the integral over $r_{\bot}^{\prime}$ in Eq.(\ref{R2_thetaScatWave}) reduce to the sum over $S$ region, that is,
\begin{equation}\label{R2_thetaScatWavev1}
	\phi_s(k_{\theta},r_{\bot}) = \sum_{r^{\prime}_{\bot}\in S} \alpha(k_{\theta},r_{\bot}^{\prime}) G^+_0(E_0; k_{\theta}, r_{\bot}-r^{\prime}_{\bot}); \,\,\,\,\,\, 
	\alpha(k_{\theta},r_{\bot}^{\prime}) = \mathcal{V}(k_{\theta},r^{\prime}_{\bot})  \phi_i(k_{\theta},r^{\prime}_{\bot}),
\end{equation}
Finally, we can expand the free Green's function in $k_{\bot}$ space and obtain the similar results as Eq.(4), that is, 
\begin{equation}\label{R2_FinalScatWave}
	\phi_s({k_{\theta}},r_{\bot}) = \sum_{r^{\prime}_{\bot}\in S} \alpha(k_{\theta},r_{\bot}^{\prime}) 
	\left\{  
	\begin{split}
		&\sum_{|z_{\mathrm{in}}|<1} {\mathrm{C}}(z_{\mathrm{in}}) z_{\mathrm{in}}^{r_{\bot}-r_{\bot}^{\prime}}, \,\,\,\,\,\,\,\,\,\,\,\,\,\,\,\,\,\, r_{\bot} > r^{R}_{\bot}; \\  
		&\sum_{|z_{\mathrm{out}}|>1} - {\mathrm{C}}(z_{\mathrm{out}}) z_{\mathrm{out}}^{r_{\bot}-r_{\bot}^{\prime}}, \,\,\,\,\,\,\,\, r_{\bot} < r^{L}_{\bot},
	\end{split}  
	\right.  
\end{equation}
where $\alpha(k_{\theta},r_{\bot}^{\prime})$ is a finite constant for each fixed $k_{\theta}$ and $r_{\bot}^{\prime}$, and $C(z)$ has the same definition as Eq.(4), that is $2\pi i$ times the residue of $[z (E_0+i\eta-H_0(k_{\theta},z))]^{-1}$. 
Note that when $\mathcal{V}(k_{\theta},r_{\bot})$ takes the delta potential along $,r_{\bot}$ direction, the Eq.(\ref{R2_FinalScatWave}) will reduce to Eq.(4) in the main text. 

From Eq.(\ref{R2_FinalScatWave}) one can see that, when dynamical degeneracy splitting occurs, for some $k_{\theta}$, there is just one pole $z$ approaching $|z|=1$, which belongs to case (ii) in the main text and results in anomalous scattering that the reflected wave are exponentially damped away from the scattering potential section $S$. 
Therefore, we have demonstrated that our results of the anomalous scattering based on Eq.(4) of the main text is robust against the perturbed scattering potential and does not depend on the special form of $V$, just requires that dynamical degeneracy splitting occurs in the free Hamiltonian. 

\begin{figure}[t]
	\begin{center}
		\includegraphics[width=.8\linewidth]{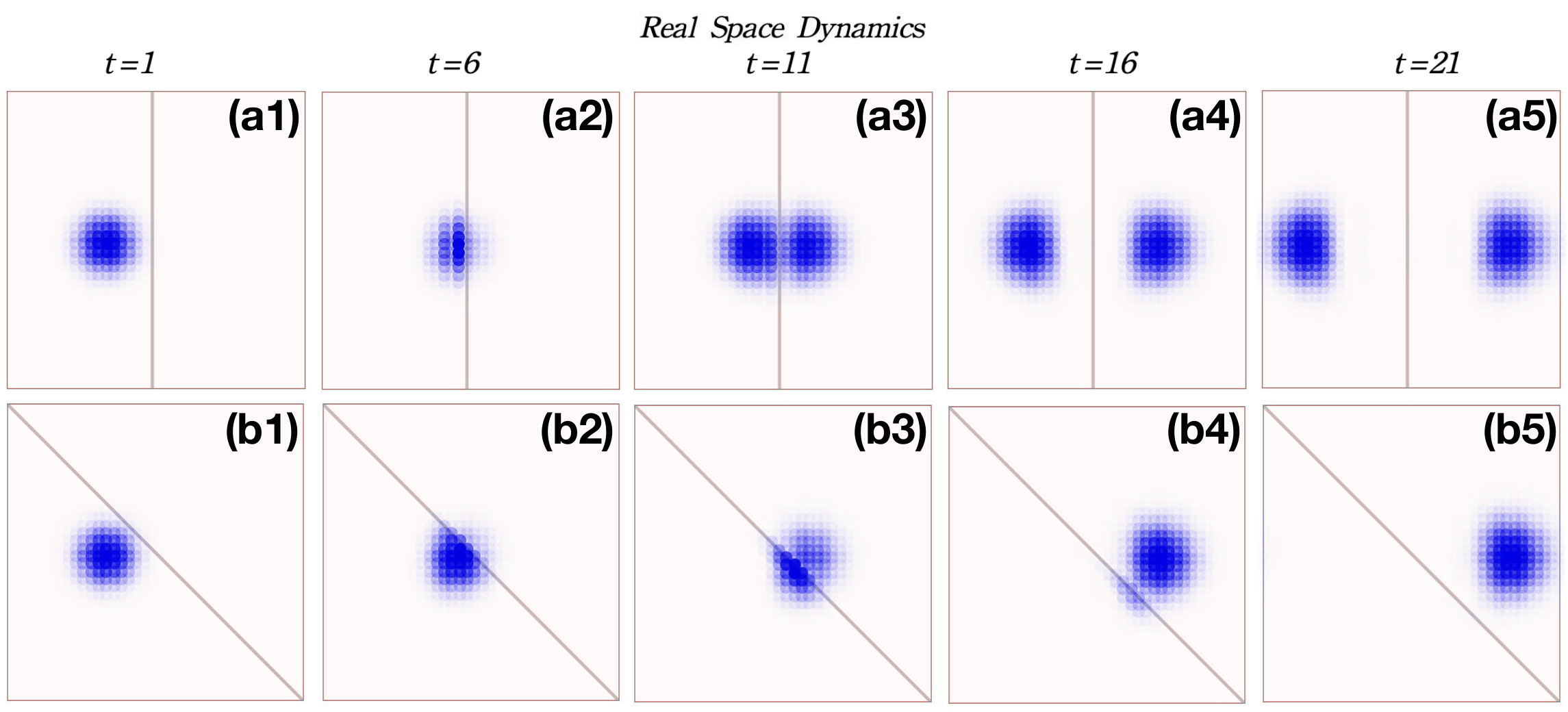}
		\par\end{center}
	\protect\caption{\label{SM_FigS6} Directional invisibility for the incident Gaussian wave packet with $(1,i)^T$ orbital components. }
\end{figure}

\subsection{The numerical simulation of anomalous scattering on finite step potential and open boundary}

Here, we present some numerical results of wave-packet scattering process on finite step potential in Fig.~\ref{SM_FigS5}(a)(b) and open boundary in Fig.~\ref{SM_FigS5}(c), which shows the similar result of anomalous scattering as in Fig.~3 of the main text. 

The free Hamiltonian of this example adopts Eq.(5) in the main text. 
The scattering potential in Fig.~\ref{SM_FigS5}(a)(b) is the step potential having the form: 
$\lambda(\bm{r})\sigma_0$, where $\lambda(\bm{r})=1$ when $\bm{r}$ is in the gray region, otherwise $\lambda(\bm{r})=0$. 
In Fig.~\ref{SM_FigS5}(c), the scattering potential is the open boundary, that is, infinite step potential. 

According to figures ~3(a)(b) in the main text, it can be seen that dynamical degeneracy splitting occurs at $\omega=3/2$. 
Therefore, we adopt the same form of incident wave packet as that in the main text, that is, $\phi_0(\bm{r}) = \mathrm{exp}[-(\bm{r}-\bm{r}_0)^2/\sigma^2+i \bm{k}_i^0 \bm{r}](1,1)^T$. 
In Fig.~\ref{SM_FigS5}(a), we can see a relatively obvious reflected wave component, while in Fig.~\ref{SM_FigS5}(b), the incident wave almost all enters the gray area, similar to the results obtained in the Fig.~3 of the main text. 
In Fig.~\ref{SM_FigS5}(c), the incoming wave packet hits the open boundary and gets stuck. 
The numerical results indicate that when dynamical degeneracy splitting occurs, the anomalous scattering based on Eq.(4) also applies to step potential and open boundary. 

\subsection{The robustness of directional invisibility}

We present the numerical results in Fig.~\ref{SM_FigS6} to verify that the directional invisibility is robust against such changes of orbital degrees of freedom. 
We adopt the Hamiltonian in Eq.(5) in the main text, and use the same incident Gaussian wave packet as that in Fig.~3 but its orbital components are changed as $(1,i)^T$. 
The real space dynamics (from $t=1$ to $t=21$) for vertical impurity line and oblique impurity line are show in Fig.~\ref{SM_FigS6}(a1)-(a5) and (b1)-(b5), respectively. 
It clearly demonstrates the directional invisibility.

\input{DDSDI_SupMat.bbl}
\end{document}

%% file: DDSDI.bbl
%